\documentclass[aps,prx,twocolumn,showpacs,superscriptaddress,footinbib]{revtex4-1}

\usepackage[english]{babel}
\usepackage{xcolor}
\usepackage{comment}
\usepackage{graphicx}
\usepackage{stackengine}
\usepackage{dcolumn}
\usepackage{bm}
\usepackage{bbm}
\usepackage[utf8]{inputenc}
\usepackage{amssymb}
\usepackage{amsmath}
\usepackage{bbold}
\usepackage{pstricks}
\usepackage{slashed}
\usepackage{braket}
\usepackage{hyperref}
\usepackage{natbib}
\usepackage{float}
\usepackage{verbatim}
\usepackage{siunitx}
\usepackage{lipsum}
\usepackage{slashed}
\usepackage{ulem}
\usepackage[left=1.85cm,right=1.85cm,top=1.85cm,bottom=1.85cm]{geometry}

\definecolor{mygreen}{rgb}{0,0.5,0}
\definecolor{myblue}{rgb}{0,0,0.75}
\definecolor{mymagenta}{cmyk}{0,1,0,0.12}
\definecolor{mygray}{rgb}{0.5,0.5,0.5}

\newcommand{\Fig}[1]{Fig.~\ref{#1}}
\newcommand{\Figure}[1]{Figure~\ref{#1}}

\begin{document}

\title{Hamiltonian Learning in Quantum Field Theories}

\author{Robert Ott}
\affiliation{Institute for Theoretical Physics, University of Innsbruck, Innsbruck, 6020, Austria}\affiliation{Institute for Quantum Optics and Quantum Information of the Austrian Academy of Sciences, Innsbruck, 6020, Austria}
\author{Torsten V. Zache}
\affiliation{Institute for Theoretical Physics, University of Innsbruck, Innsbruck, 6020, Austria}
\affiliation{Institute for Quantum Optics and Quantum Information of the Austrian Academy of Sciences, Innsbruck, 6020, Austria}

\author{Maximilian Pr\"{u}fer}
\affiliation{Vienna Center for Quantum Science and Technology,
Atominstitut, TU Wien, Stadionallee 2, 1020 Vienna, Austria}

\author{Sebastian Erne}
\affiliation{Vienna Center for Quantum Science and Technology,
Atominstitut, TU Wien, Stadionallee 2, 1020 Vienna, Austria}

\author{Mohammadamin Tajik}
\affiliation{Vienna Center for Quantum Science and Technology,
Atominstitut, TU Wien, Stadionallee 2, 1020 Vienna, Austria}

\author{Hannes Pichler}
\affiliation{Institute for Theoretical Physics, University of Innsbruck, Innsbruck, 6020, Austria}\affiliation{Institute for Quantum Optics and Quantum Information of the Austrian Academy of Sciences, Innsbruck, 6020, Austria}

\author{J\"{o}rg Schmiedmayer}
\affiliation{Vienna Center for Quantum Science and Technology, Atominstitut, TU Wien, Stadionallee 2, 1020 Vienna, Austria}

\author{Peter Zoller}
\affiliation{Institute for Theoretical Physics, University of Innsbruck, Innsbruck, 6020, Austria}
\affiliation{Institute for Quantum Optics and Quantum Information of the Austrian Academy of Sciences, Innsbruck, 6020, Austria}

\begin{abstract}
We discuss Hamiltonian learning in quantum field theories as a protocol for systematically extracting the operator content and coupling constants of effective field theory Hamiltonians from experimental data. Learning the Hamiltonian for varying spatial measurement resolutions gives access to field theories at different energy scales, and allows to learn a flow of Hamiltonians reminiscent of the renormalization group. Our method, which we demonstrate in both theoretical studies and available data from a quantum gas experiment, promises new ways of addressing the emergence of quantum field theories in quantum simulation experiments.
\end{abstract}
\maketitle

\section{Introduction}
Quantum simulation aims at predicting the properties of isolated quantum many-body systems for a model Hamiltonian of interest through engineered quantum devices~\cite{national2020manipulating,gross2023physics}. Hamiltonian Learning (HL)~\cite{daley2022practical,carrasco2021theoretical,eisert2020quantum,anshu_sample-efficient_2021,qi2019determining,bairey_learning_2019,li_hamiltonian_2020,wiebe2014hamiltonian,valenti2022scalable,pastori2022characterization,shabani2011estimation} addresses the inverse problem, specifically, the acquisition of the \emph{operator content} of the system Hamiltonian $\hat{H}_{\text{sys}}$. This Hamiltonian is either a priori unknown or imprecisely known. The learning process is based on a protocol which reconstructs the Hamiltonian from experimental data obtained through a quantum simulator.

For lattice models, Hamiltonian learning has been well-established, and learning protocols have been explored for systems prepared in equilibrium states~\cite{anshu_sample-efficient_2021,bairey_learning_2019,qi2019determining} and for non-equilibrium quench dynamics~\cite{bairey_learning_2019,li_hamiltonian_2020,wiebe2014hamiltonian,shabani2011estimation,pastori2022characterization,valenti2022scalable}. These protocols enable efficient inference of the operator structure of $\hat{H}_{\text{sys}}$ from measurements. Experimental setups encompass the Bose or Fermi Hubbard models with ultracold atoms in optical lattices, as well as spin models implemented with Rydberg tweezer arrays, trapped ions, or superconducting circuits~\cite{national2020manipulating,browaeys2020many,altman2021quantum}. In these setups, single-shot, single-site resolved measurements via a quantum gas microscope~\cite{kuhr2016quantum} or single-shot readout of individual spins or qubits~\cite{altman2021quantum} provide the required experimental data for Hamiltonian reconstruction, including an assessment of errors.

\begin{figure}[t!]
	\includegraphics[width=1.\columnwidth]{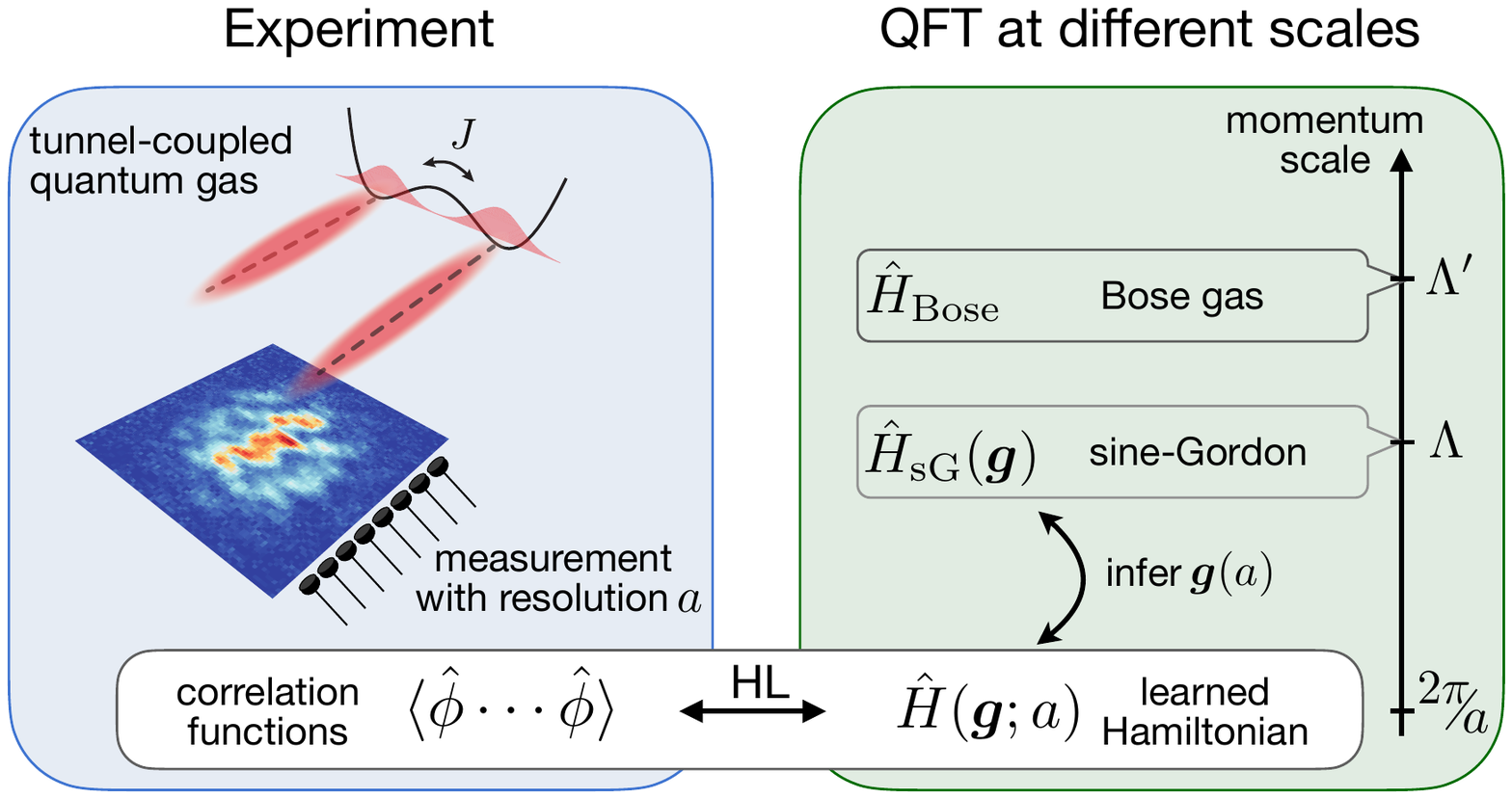}
	\caption{\textit{Hamiltonian learning in QFT.}
 (Left) In ultracold quantum gas experiments, measurements with a resolution scale $a$ allow to determine the coarse-grained correlation functions of fields $\hat{\phi}_n$ and their canonical momenta $\hat{\pi}_n$ (and combinations thereof). Here, we exemplify this with a measurement of field correlations through interference patterns of two tunnel-coupled one-dimensional Bose gases. (Right) QFTs are valid below a cutoff scale $\Lambda'$, where for instance a cloud of tunnel-coupled ($J$) atoms can be described by a Bose gas Hamiltonian~$\hat{H}_\mathrm{Bose}$ with contact interactions, independent of subatomic details. At even larger distance scales a description in terms of an effective sine-Gordon field theory with Hamiltonian $\hat{H}_\mathrm{sG}(\boldsymbol{g})$~\eqref{eq:Hscalar} and cutoff $\Lambda$ may emerge.
 To develop HL for QFTs, we match the involved scales by constructing an effective, coarse-grained ansatz~$\hat{H}(\boldsymbol{g};a)$~\eqref{eq:ansatz}, which can be learned from the experimentally observed correlation functions at the resolution scale~$a$.
 }
	\label{fig:Scheme0}
\end{figure}

Here, we develop Hamiltonian learning for quantum field theories (QFTs), which represent theories of interacting quantum fields formulated in continuous space, a conceptual framework familiar from particle physics~\cite{weinberg1995quantum} and condensed matter physics~\cite{altland2010condensed,sachdev2023quantum,fradkin2013field}. This includes the experimentally relevant examples
of interacting quantum degenerate Bose and
Fermi gases of ultracold atoms~\cite{national2020manipulating,gardiner2017quantum}, where quantum fields are associated
with point-like atoms as basic constituents of the theory. QFTs must be understood as an effective description valid in a given energy range defined by a cutoff $\Lambda$, and as a function of the cutoff a family of Hamiltonians emerge, which are connected by the renormalization group (RG) flow~\cite{fisher1998renormalization}. We sketch QFTs on different scales in the right panel of~\Fig{fig:Scheme0}.\\[0.2cm]

An example from ultracold atom experiments is given by tunnel-coupled one-dimensional Bose gases (see left panel of \Fig{fig:Scheme0}) \cite{schweigler2017experimental,pigneur2018relaxation,tajik2023verification}. While a microscopic description of a quantum degenerate Bose gas might start at the level of weakly interacting bosonic atoms with an effective Hamiltonian $\hat{H}_\mathrm{Bose}$ involving a point-like pseudo-potential interaction with scattering length $a_s$~\cite{gardiner2017quantum}, on a more coarse-grained scale the physics is modeled by an emergent sine-Gordon QFT \cite{gritsev2007linear,dalla2013universal,imambekov2008mapping} with Hamiltonian~\cite{giamarchi2003quantum,cuevas2014sine}
\begin{equation}
\label{eq:Hscalar}
    \hat{H}_{sG} = \int_\Lambda  \mathrm{d}x\, \Big[\frac{1}{2}\hat{\pi}_x^2 + \frac{1}{2}(\nabla\hat{\varphi}_x)^2  -g \cos(\hat{\varphi}_x) \Big]\, .
\end{equation}
This Hamiltonian represents a paradigmatic, relativistic QFT relevant in the context of condensed matter physics~\cite{kogut1979introduction}, gauge theories~\cite{coleman1976more} and cosmology~\cite{cuevas2014sine}, where $\hat{\varphi}_x$ and $\hat{\pi}_x$ are the continuous quantum fields and their conjugate momenta. In cold-atom experiments, they describe the phase and density fluctuations of the atoms controlled by the interaction term with strength~$g$~\cite{gritsev2007linear}.

Observables accessible in experiments are extracted with various imaging techniques; in real space, for example, using quantum gas microscopes, or in momentum space after a free expansion of the gas. In both cases every atom can be detected with close to unit efficiency~\cite{Buecker_2009,bakr2009quantum}. The measurements are performed with a finite resolution~$a$, and thus, these experiments measure a hierarchy of coarse-grained correlation functions $\langle \hat{\phi}\cdots\hat{\phi} \rangle$ (see Sec.~\ref{sec:HL-theory}), as for example studied experimentally in~Refs.~\cite{schweigler2017experimental,langen2015experimental,Feng2019,zache2020extracting,prufer2020experimental,schweigler2021decay,tajik2023verification}.

HL reconstructs the effective Hamiltonian including the operator content and coupling constants $\boldsymbol{g}$ underlying all these coarse-grained correlation functions at the given resolution scale $a$. This is indicated in \Fig{fig:Scheme0} as the learned Hamiltonian $\hat{H}(\boldsymbol{g};a)$ (defined in Eq.~\eqref{eq:ansatz}). Changing the resolution scale $a$ implies a flow of the learned couplings $\boldsymbol{g}\equiv\boldsymbol{g}(a)$ reminiscent of the flowing couplings of the RG (see Sec.~\ref{sec:RG-flow}). This flow can be extrapolated to confront this learned coarse-grained Hamiltonian from experimental data with theoretical models of a continuum-effective field theory. Examples we consider in this work include the free field theory quadratic in operators $\hat{\varphi}$, as well as interacting QFTs with quartic operators $\hat{\varphi}^4$, or the sine-Gordon field theory~\eqref{eq:Hscalar}, where we distinguish different ansätze for the learned Hamiltonians with an error assessment based on constraints for the measured correlation functions. We  illustrate this below both with numerical simulations and available experimental data~\cite{schweigler2017experimental}.

While this paper focuses on HL in QFTs for specific examples, which includes the limit of a classical-statistical equilibrium state, we stress that the methods developed here can be applied in a broader context. Furthermore, we detail the requirements in terms of resolution and measurement budget for HL for both thermal states across a broad range of temperatures, and time-evolved systems after quenches.

The paper is structured as follows: We discuss the theoretical concepts and the workflow behind HL in QFTs in Sec.~\ref{sec:HL-theory}. There, we illustrate HL in different relevant settings in- and out of thermal equilibrium. In Sec.~\ref{sec:application} and Sec.~\ref{sec:Experiment}, we first apply HL to numerical simulations of the sine-Gordon model in the classical-statistical high-temperature regime, and subsequently demonstrate HL with experimental data of the experiment detailed in~\cite{schweigler2017experimental}. The dependence of HL on the measurement budget is further discussed in Sec.~\ref{sec:statistics}, before we close with conclusions in Sec.~\ref{sec:conclusion}.

\section{Hamiltonian learning in QFTs}
\label{sec:HL-theory}
In this section, we discuss the ideas behind Hamiltonian learning for QFTs and illustrate its implications for an explicit example of a one-dimensional QFT. To this end, we simulate a quantum simulation experiment similar to the one shown in \Fig{fig:Scheme0}, by computing the ideal measurement outcome in the form of field correlation functions analytically. In the present case, we generate these correlations from the Hamiltonian 
\begin{align}
	\label{eq:H-sys}
	\hat{H}_\mathrm{sys}= \int_\Lambda  \mathrm{d}x\, \Big[\frac{1}{2}\hat{\pi}_x^2 + \frac{1}{2}(\nabla\hat{\varphi}_x)^2 +\frac{ m^2}{2} \hat{\varphi}_x^2 +\lambda\hat{\varphi}^4_x\Big]\, ,
\end{align}
which involves the two coupling parameters of the mass~$m$, and the interaction strength $\lambda$, whose values are set by the underlying details of the experimental setup. The subscript $\Lambda$ denotes the cutoff scale below which the field theory is expected to become a valid description of the (simulated) experimental data. In an experiment, such a scale would again be set by the underlying microscopic ingredients, e.g. the scattering length of atoms in a cold quantum gas. To mimic such a scale in our simulations, we either implement the cutoff via a lattice regularization with lattice spacing $a_0$ (such that $\Lambda=\pi/a_0$), or with a hard cutoff for momentum modes. In what follows, we describe the learning procedure assuming that we have no detailed knowledge about the operator structure of $\hat{H}_\mathrm{sys}$, nor about the value of the coupling constants.

As indicated in \Figure{fig:Scheme0}, the simulated measurement outcome is given by time-local (``equal-time'') snapshots of coarse-grained field operators 
\begin{equation}
	\label{eq:coarse-data}
	\hat{\phi}_{n} = \int \mathrm{d}x\, W_{nx} \hat{\varphi}_{x} \, ,
\end{equation}
along with the corresponding canonical momenta $\hat{\pi}_n$. Here, $W$ describes the reduction from the continuous quantum fields $\varphi_x$, to the discrete set of modes $\phi_n$ that are observable by the measurement apparatus, labeled by integers $n$. In practice, $W$ is set by the experimental imaging system and explicitly depends on the resolution scale $a$. This (non-invertible) reduction from continuous to discrete indices represents the loss of information by discarding the short-distance modes in the measurements. In this section, we consider the simple model $W_{nx} = \mathrm{sinc}((x-na)/\sigma)$, exemplifying measurements with optical resolution $\sigma = a$ at positions~$na$. This choice introduces a hard cutoff in momentum space, where field modes corresponding to distance scales smaller than $a$ are discarded. While such a choice simplifies the explicit calculations in this section, the main messages of our results do not depend on it. In the numerical and experimental section further below, different, more realistic measurement models are used. 

Repeated projective measurements then allow access to spatially dependent correlation functions of these coarse-grained fields, $\langle \hat{\phi}_n \hat{\phi}_m \cdots \rangle$, as well as the canonical momenta  $\langle \hat{\pi}_n \hat{\pi}_m \cdots \rangle$, which represent the experimental input to our learning protocol.

The Hamiltonian is then learned in a step of classical post-processing, where different such correlation functions are related to each other by constraint equations which follow from a Hamiltonian ansatz. As indicated on the right-hand side of \Fig{fig:Scheme0}, we first formulate a field-theoretic ansatz \begin{align}
	\label{eq:ansatz}
	\hat{H}(\boldsymbol{g}) = \int_\Lambda  \mathrm{d}x\, \Big[\frac{1}{2}\hat{\pi}_x^2 + \frac{1}{2}(\nabla\hat{\varphi}_x)^2 + V(\boldsymbol{g};\hat{\varphi}_x)\Big]\, ,
\end{align} at the cutoff scale $\Lambda$ including a set of (ansatz) coupling constants $\boldsymbol{g}=\{g_i \,| i \in \mathbf{N}\}$ to be learned and the potential
\begin{align}
	V(\boldsymbol{g};\hat{\varphi}_x) = \sum_{i\in  \mathbf{N}} g_i \hat{\varphi}_x^i\, .
\end{align} 
 Starting from this microscopic ansatz, we analytically derive a
 ``coarse-grained'' ansatz Hamiltonian $\hat{H}(\boldsymbol{g};a)$ (see \eqref{eq:ansatz} in section \ref{sec:eff-H}) which acts on the resolution scale $a$ and which is formulated in terms of the experimentally accessible field operators $\phi_n$. That is correlations involving this Hamiltonian, e.g. $ \langle \hat{H}(\boldsymbol{g};a) \rangle $, can be readily accessed using the measurement outcome described above. The derivation of such an effective Hamiltonian can be achieved by integrating out the unobservable field modes from the path integral as detailed in the next section \ref{sec:eff-H}.

The Hamiltonian $\hat{H}(\boldsymbol{g};a)$ then implies the constraint equations $\mathcal{C}(\boldsymbol{g}) = 0$, which explicitly depend on the resolution scale $a$ and the variational coupling parameters~$\boldsymbol{g}$. In the simplest case, they correspond to the conservation of the effective Hamiltonian during real-time dynamics from various initial states, or the stationary state condition for equilibrium states. Optimizing the constraints by finding the best couplings $\boldsymbol{g}$ then allows us to learn the system Hamiltonian~$\hat{H}_\mathrm{sys}$. The Hamiltonian is learned successfully if a set of coupling constants $\boldsymbol{g}$ is found such that the constraint equations are fulfilled within the desired precision $\mathcal{C}(\boldsymbol{g}) \approx 0$. In turn, if this is not possible, i.e. $\mathcal{C}(\boldsymbol{g}) \neq 0$, Hamiltonian learning with the given ansatz fails and hence updating the ansatz is necessary, that is, adding additional interaction terms, $\hat{H}(\boldsymbol{g})\rightarrow \hat{H}(\boldsymbol{g}')$. In general, the final value of the constraint gives us an estimator for the correct implementation of the field theory and allows us to prefer Hamiltonians $\hat{H}(\boldsymbol{g}')$ over $\hat{H}(\boldsymbol{g})$ when $\mathcal{C}(\boldsymbol{g}')$ is better fulfilled than~$\mathcal{C}(\boldsymbol{g})$.

Throughout this section, we demonstrate Hamiltonian learning for quantum field theories in different settings. Starting from the ansatz $\hat{H}(\boldsymbol{g})$, we first show how we construct the effective Hamiltonian $\hat{H}(\boldsymbol{g};a)$. Next, we illustrate Hamiltonian learning from real-time measurements, after performing mass quenches in a free ($\lambda=0$) quantum field theory. Furthermore, we discuss Hamiltonian learning for thermal equilibrium states, in which context we observe a resolution-scale flow of the learned coupling constants in the interacting field theory. At last, we generalize this procedure to thermal states at high temperatures, which is relevant for the following numerical and experimental semi-classical demonstrations.
Throughout this section, our protocol is illustrated for an idealized scenario where we use analytical calculations to ``simulate" a perfect experimental realization of the field theory. A more detailed analysis for a finite measurement budget for the statistical sample size follows in the later section \ref{sec:statistics} below.

\subsection{Effective Hamiltonian ansatz}
\label{sec:eff-H}

To construct the effective Hamiltonian $\hat{H}(\boldsymbol{g};a)$, we employ a path integral formalism. The path integral allows us to efficiently parametrize the observable correlation functions in terms of source fields $J$ and to implement the loss of information due to a finite resolution.

We separate the action of the quantum fields into observable and inobservable fields denoted by~$\phi$ and~$\eta$. This separation is motivated by the measurement. In the current model $W_{nx}$, the measurement introduces a new cutoff $\Lambda_a=2\pi/a$ in momentum space, i.e. observable fields~$\phi$ are those with momenta $p<\Lambda_a$, and inobservable fields $\eta$ live on larger momentum scales $p>\Lambda_a$, which are discarded in the measurement.

Correlation functions can be obtained from functional derivatives (with respect to $J_n$) of the path integral. The Hamiltonian of the system is encoded in the vacuum path integral 
\begin{equation}
    Z[J] = \int \mathcal{D}\varphi \, e^{iS[\boldsymbol{g};\varphi]+\sum_{n}J_n\phi_n}\, ,
\end{equation}
where the integral measure is defined as $\mathcal{D}\varphi =\prod_{p<\Lambda}\mathrm{d}\varphi_p$, and source fields only couple to the observable field variables~$\phi_n$, thus encoding the finite resolution. $S$ is the action corresponding to the Hamiltonian \eqref{eq:ansatz},
\begin{equation}
S[\boldsymbol{g};\varphi] =\int \mathrm{d}\tau \int \mathrm{d}x \left[\frac{1}{2} (\partial_\mu \varphi)^2 - \frac{m^2}{2} \varphi^2 + V(\boldsymbol{g};\varphi)\right]\, , \end{equation}
with $\partial_\mu$ being the partial derivative with respect to coordinate $\mu = \tau, x$ for temporal and spatial directions. 
The path integral measure can be split into observable and inobservable modes as $\mathcal{D}\varphi = \mathcal{D}\phi \mathcal{D}\eta$.

In the following, we assume that the fields $\eta$ and $\varphi$ are weakly interacting, such that they approximately decouple.
Integrating out the fields~$\eta$ then leads to the effective action
\begin{align}
S^\mathrm{eff}[\boldsymbol{g};a;\phi]\! =\! \int \mathrm{d}\tau  \Big[\sum_{n}\frac{1}{2}\dot{\phi}^2_n &+ \sum_{n,m}\frac{1}{2}\phi_m M_{mn} \phi_n  \nonumber\\
&+ V(\boldsymbol{g},\phi_n) \Big]\, ,
\end{align}
which explicitly depends on the resolution scale $a$. Here, the ``mass'' matrix is obtained as
\begin{align}
M_{mn} = \left(\frac{\pi^2}{a^3}+ \frac{m^2}{a} \right)\delta_{mn} +\frac{2}{a^3}\frac{(-1)^{n-m}}{(n-m)^2} \, .
\end{align}
As detailed in the appendix \ref{sec:scale-dependent-quantum}, $M_{mn}$ represents the coarse-grained analog of the differential operator $-\partial^2+m^2$ without further corrections from the interactions. In summary, we obtain the effective, resolution scale-dependent Hamiltonian
\begin{align}
\label{eq:ansatz}
    \hat{H}(\boldsymbol{g};a) = \frac{a}{2}\sum_n \hat{\pi}_n^2+ \frac{a^2}{2}\sum_{n,m} \hat{\phi}_n M_{nm}\hat{\phi}_m +  a\sum_n V(\boldsymbol{g};\hat{\phi}_n) \, ,
\end{align}
where the coarse-grained conjugate fields for our choice of $W$ are given by $\hat{\pi}_n = \int \mathrm{d}x \, W_{nx} \hat{\pi}_x$, such that $[\hat{\phi}_n,\hat{\pi}_m] = i\delta_{nm}$. While Eq.~\eqref{eq:ansatz} takes the form of a lattice Hamiltonian, we emphasize that here the measurement resolution sets the discretization of the spatial coordinate.

\begin{figure}[t!]
	\includegraphics[width=1.03\columnwidth]{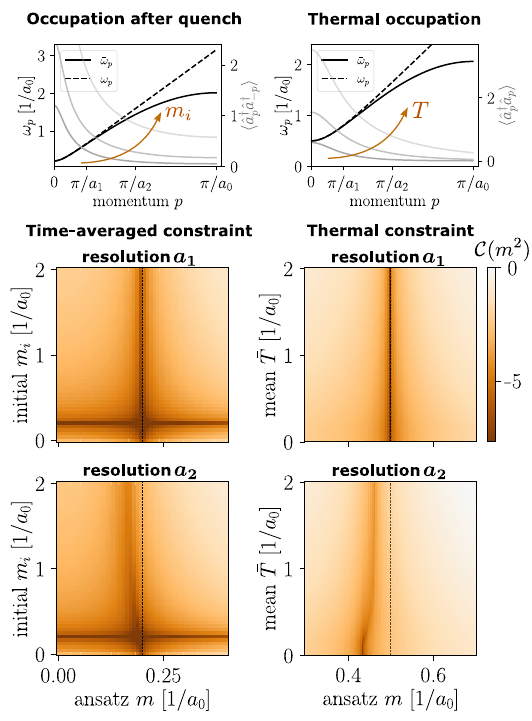}
	\caption{\textit{Learning the Hamiltonian from time-evolved and thermal free quantum fields (exact analytics)}. We illustrate Hamiltonian learning for two scenarios: from the time-evolution of the quantum system after a quench $m_i\rightarrow m=0.2/a_{0}$ (left), and the field correlations in thermal equilibrium states with~$ m=0.5/a_{0}$ (right). In the top row, we show lattice ($\bar{\omega}_p$) and continuum ($\omega_p$) dispersion relations (with a.u.), which agree in the IR, where $p\ll \pi/a_0$. Different quench strengths and temperatures lead to varying occupations of UV modes where continuum and lattice differ. Below, we show the constraint landscape for two resolutions (plotted logarithmically): $a_1=0.2 \pi/a_{0}$ (top) and $a_2=0.5 \pi/a_{0}$ (bottom). The result of our learning scheme is given by the minima of the plots shown here. (Left, $a_1$) For non-zero quenches, the minimum is close to the ``correct'' mass (dashed line) for all quenches, as only IR modes are resolved. (Left, $a_2$) With a better resolution, the difference of lattice and continuum theories is resolved once UV modes are occupied for sufficiently strong quenches. As a consequence, the constraint cannot be fulfilled. (Right)~A similar qualitative picture is obtained in the thermal equilibrium case. Here, the mode occupation is set by the thermal Bose-Einstein distribution. In contrast to the real-time case, we have a non-zero vacuum contribution (``quantum half") at all momenta, such that for $a=a_2$ a non-zero shift appears even for small temperatures. 
 }
	\label{fig:real-time}
\end{figure}
\subsection{Learning the Hamiltonian from real-time dynamics}
As the Hamiltonian generates time translations, it can be learned using time-evolved observables of the system~\cite{li_hamiltonian_2020}. Specifically, this implies that the correct Hamiltonian is a constant of motion during the time-evolution of quantum states, i.e. $\bra{\psi(t)} [\hat{H}(\boldsymbol{g};a)]^n\ket{\psi(t)} \approx \bra{\psi(0)} [\hat{H}(\boldsymbol{g};a)]^n\ket{\psi(0)} $ for any time $t$ and integer $n$, if the system Hamiltonian is well-described by $\hat{H}(\boldsymbol{g};a)$.  In the simplest case of $n=1$, this yields the constraint $ \mathcal{C}_t(\boldsymbol{g}) = 0$, with
\begin{align}
		\mathcal{C}_t(\boldsymbol{g}) &=  \frac{a}{2}\sum_n [\langle\hat{\pi}_n^2\rangle_t-\langle\hat{\pi}_n^2\rangle_0]\nonumber\\
		&+ \frac{a^2}{2}\sum_{n,m} M_{nm}[\langle\hat{\phi}_n \hat{\phi}_m\rangle_t-\langle\hat{\phi}_n \hat{\phi}_m\rangle_0] \nonumber\\
		&+  a \sum_n[\langle V(\boldsymbol{g};\hat{\phi}_n)\rangle_t -\langle V(\boldsymbol{g};\hat{\phi}_n)\rangle_0 ]\, ,
\end{align}
where $\langle\cdot\rangle_t = \bra{\psi(t)}\cdot\ket{\psi(t)}$ refers to the time-dependent expectation value. This constraint depends explicitly on the coupling parameters $g_i$ whose value we aim to determine in Hamiltonian learning.

As an example, we consider the dynamics of a free ($\lambda=0$) scalar field theory on a UV lattice with spacing $a_{0}$ after quenches of the bare mass. The implementation on a lattice explicitly sets an underlying microscopic scale, as also present in every experiment where quantum fields emerge effectively, and it can be interpreted as a specific choice of a UV regularization of the system Hamiltonian $\hat{H}_\mathrm{sys}$ in Eq.~\eqref{eq:H-sys}. Here, we expect a description of the correlations in terms of the continuum field theory towards the infrared, while deviations are expected once the details of this UV regularization are resolved at the lattice scale. For the free theory considered here, the physics is captured by the free dispersion relation, which is given by $\omega^2_p = m^2+p^2$ in the continuum theory, but it takes the form $\bar{\omega}^2_p = m^2+(2/a_0)^2\sin^2(pa_0/2)$ on the discrete lattice. While they agree for momenta $p\ll \pi/a_0$, strong differences occur close to the lattice scale, see \Fig{fig:real-time}.
 
 We compute the dynamics of the correlation functions using the lattice system Hamiltonian with mass $m=0.2/a_{0}$, but starting from ground states corresponding to the masses $m_i$ (all masses are expressed in terms of the UV lattice constant $a_0$). The different quenches lead to a time-evolution of the two-point functions depending on the ``occupation" $\langle \hat{a}^\dagger_p \hat{a}^\dagger_{-p}\rangle$ created after the quench, where $\hat{a}^\dagger_p$ is the field creation operator for momentum $p$, see the appendix~\ref{sec:quench} for details. Different quench ``strengths" are illustrated in \Fig{fig:real-time} for varying~$m_i$: For weak quenches, i.e. $|m_i-m|\ll m$, the initial state mostly occupies infrared modes, while the occupied modes extend further to the UV for stronger quenches.

 We demonstrate the HL protocol for two resolutions $a_1= 0.2 \pi/a_0$ and $a_2= 0.5 \pi/a_0$. Importantly, we assume that the Hamiltonian is the same one throughout the considered evolution times. We thus combine constraints at different evolution times $t$, i.e. $\mathcal{C}(m^2) = (1/N_t)\sum_t\,  |\mathcal{C}_t(m^2)|$ which we demand to be fulfilled simultaneously (in \Fig{fig:real-time} we average over three times to approximate the time average). In the first case, $a=a_1$, continuum and lattice theory agree to a good degree for all observed momentum scales $p<\pi/a$, independent of the occupation~$\langle \hat{a}^\dagger_p \hat{a}^\dagger_{-p}\rangle$. Hence, for non-zero quenches, $m_i\neq m$, HL yields the correct value of the mass for any quench strength. Conversely, for $a=a_2$ there are considerable differences between the continuum and lattice theories. Weak quenches lead to infrared contributions such that the field theory Hamiltonian remains a good description and the correct mass can be learned. However, the difference between lattice and continuum field theory is enhanced by mode occupations in the UV region which occur for strong quenches. As a consequence, the constraints become less well fulfilled for any choice of the ansatz mass.
 
Some of the features presented here are specific to the free theory. In general, for interacting theories, one would observe particle and energy exchange among different momenta. Once momenta on the order of the resolution scale are occupied, this would lead to a failure of HL as some of the energy would be transferred to unobservable modes such that the energy conservation of observable modes is violated. We will investigate the effects of interactions on HL further below in an equilibrium setting.

\subsection{Learning the Hamiltonian from equilibrium states}
The above discussion can be extended to the case of stationary states, as exemplified in this section for thermal equilibrium~\cite{anshu_sample-efficient_2021,bairey_learning_2019}, i.e. for states $\hat{\rho}_\beta \propto e^{-\beta\hat{H}_\mathrm{sys}}$.  Again, the effective Hamiltonian~\eqref{eq:ansatz} implies constraint equations for equal-time correlation functions. These may be obtained through path-integral methods using Dyson-Schwinger equations. Alternatively, we may use the operator algebra of scalar fields $[\hat{\phi}_n,\hat{\pi}_m] = i\delta_{nm}$ and exploit the stationary state condition, i.e. $\mathrm{Tr}(\hat{\rho}_\beta [\hat{\mathcal{O}},\hat{H}(\boldsymbol{g})]) = 0$. One of the simplest of these constraints, using $\hat{\mathcal{O}}= \frac{1}{2}\{\hat{\pi}_m ,\hat{\phi}_n\}$ with anti-commutator $\{\cdot\}$, reads $\mathcal{C}_\beta(\boldsymbol{g}) = 0$, with
\begin{align}
	\label{eq:int-q-constraint}
	\mathcal{C}_\beta(\boldsymbol{g}) =  \langle\hat{\pi}_{n}\hat{\pi}_{m}\rangle_\beta + a\sum_l M_{ml} \langle\hat{\phi}_{n}\hat{\phi}_{l}\rangle_\beta +\langle\hat{\phi}_{n} V'(\boldsymbol{g};\hat{\phi}_m)\rangle_\beta \, ,
\end{align}
where $V'$ refers to the derivative of the potential with respect to the fields and $\langle \cdot\rangle_\beta$ is the thermal expectation value. 

In general, arbitrarily many of such constraints may be constructed by choosing different observables of $\hat{\mathcal{O}}$. For example, observables of the form $\hat{\mathcal{O}}= \{\hat{\pi}_m, f[\hat{\phi}]\}$ probe the Hamiltonian on higher-order field correlation functions based on (arbitrary) field functionals $f$, which corresponds to the equal-time limit of the Schwinger-Dyson equations. Typically, constraints at equal times involve both field correlations as well as cross-correlations of fields with their canonical momenta. It is important to note that there is a substantial amount of freedom in choosing the constraints. For constraints to be restrictive for different Hamiltonian couplings, they need to be sensitive to different aspects of the underlying theory. The idea is to choose constraints that involve a representative set of correlation functions, which are also relevant for observables of interest in the learned field theory. In appendix \ref{app:choice-of-constraints} we describe how different constraints are sensitive to different terms in the Hamiltonian, e.g. how constraints are chosen to efficiently detect symmetry-violating interactions. 

Furthermore, different scales of the quantum system can be probed by considering different spatial points $n,m$ of the constraint equations. Here, we focus on the constraint involving all momentum scales up to the resolution scale which determines the spacing of the spatial points, that is, we consider the constraint $\mathcal{C}_\beta(\boldsymbol{g}) =  \sum_m[\langle\hat{\pi}^2_{m}\rangle_\beta + a\sum_l M_{ml} \langle\hat{\phi}_{m}\hat{\phi}_{l}\rangle_\beta +\langle\hat{\phi}_{m} V'(\hat{\phi}_m)\rangle_\beta]$. We furthermore again combine constraints from different temperatures~$T$ to be fulfilled simultaneously (here, $\mathcal{C} =|\mathcal{C}_\beta(\boldsymbol{g})|+|\mathcal{C}_{\beta=\infty}(\boldsymbol{g})|$ with mean temperature $\bar{T} = T/2$) under the assumption that the same Hamiltonian is valid for all temperatures within the considered range. As discussed in the previous section, we expect a continuum description of the data for resolutions and temperatures in the infrared. 

In \Fig{fig:real-time}, we consider an example, where we generate correlation functions from a thermal state of a free lattice Hamiltonian. We learn the Hamiltonian couplings for a free continuum field theory Hamiltonian in analogy to the previous section. Again, we find the correct mass parameter across all mean temperatures $\bar{T}$ for resolution $a=a_1$, which probes the Hamiltonian for the infrared momenta $p<\pi/a_1 \ll \pi/a_0$.  In the second case with resolution $a=a_2$, details of the lattice scale are resolved. At small temperatures, we learn the mass parameter with a finite shift from the ``quantum half'', i.e. the vacuum fluctuations across all observed
scales including UV momenta with $p\sim\pi/a_2$. At higher temperatures the ansatz fails and no good minimum of $|\mathcal{C}|$ can be found. This is because the thermal occupation of UV modes yields a different mass shift compared to the vacuum quantum fluctuations and hence there is no single continuum Hamiltonian to describe both cases.

\begin{figure*}[t!]
	\includegraphics[width=1.8\columnwidth]{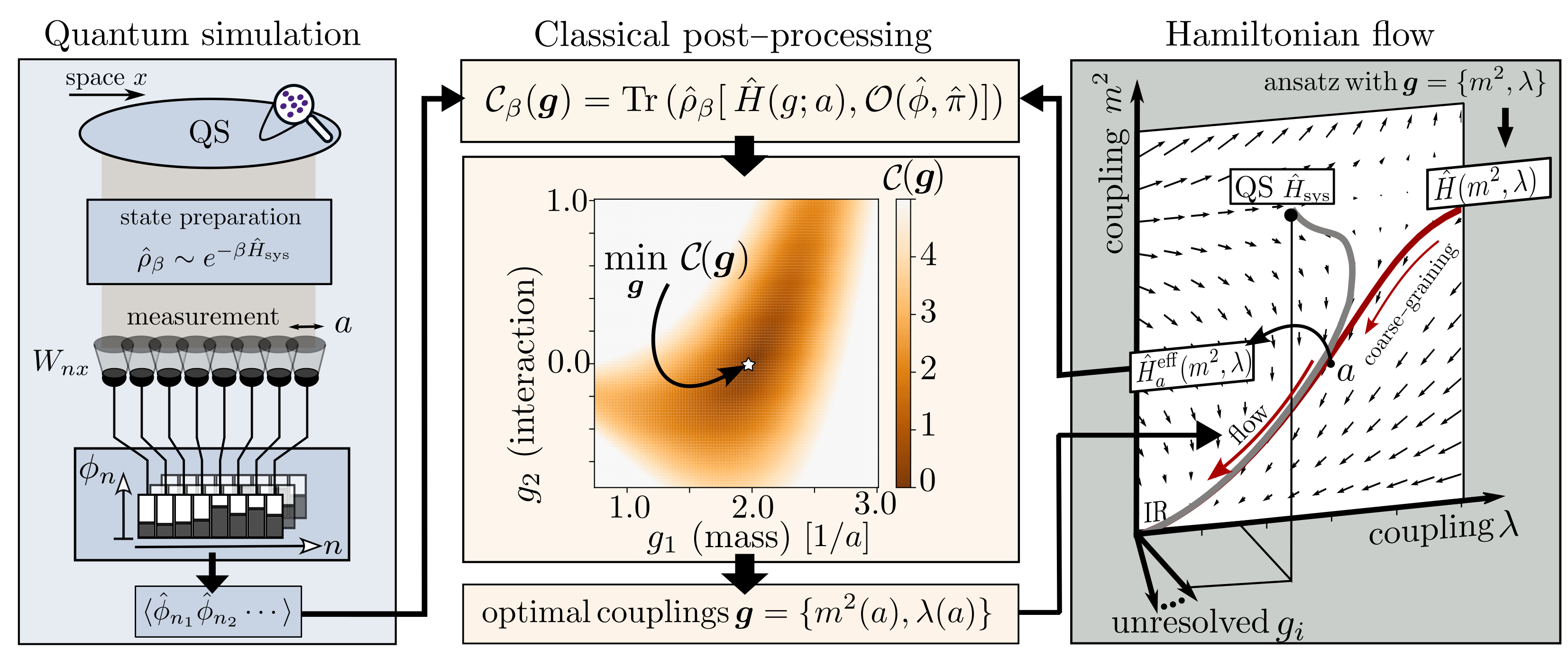}
	\caption{\textit{Summary of Hamiltonian learning workflow.}~ The goal is to learn an effective Hamiltonian for a description of a quantum simulation in terms of a quantum field theory. (Left) We consider a quantum system (QS) with spatial degrees of freedom; this can be for example a gas of ultracold atoms (violet dots). After state preparation one performs a measurement introducing a measurement scale~$a$, for example, given by the imaging resolution or the pixel size of a detector. From repeated measurements with spatial resolution, one can infer expectation values of correlation functions $\langle \hat{\phi}_n \hat{\phi}_m \cdots \rangle$. (Right) On the other side, we choose an ansatz quantum field theory Hamiltonian $\hat{H}(\boldsymbol{g})$, which we constrain to the manifold with $\boldsymbol{g}=\{m^2,\lambda\}$ for the example of a scalar $\lambda\varphi^4$ theory. The coarse-graining procedure (see main text) leads to an effective Hamiltonian $\hat{H}(m^2,\lambda;a)$ which depends on the resolution scale $a$. From this effective Hamiltonian we calculate constraints into which we feed the correlation functions obtained from the quantum simulation (Center). Optimization of the constraints by varying the couplings leads to the optimal learning results, as exemplified here for a free quantum field theory with interacting ansatz Hamiltonian, and a constraint $\mathcal{C}$ combined from two temperatures.}
	\label{fig:Scheme1}
\end{figure*}

In the following, to illustrate Hamiltonian learning for interacting fields, we focus on the continuum limit and generate the correlation functions from the continuum Hamiltonian \eqref{eq:H-sys} for simplicity. As a first step, we again consider free correlations with $\lambda = 0$ but include a $\lambda\varphi^4$ interaction in the ansatz Hamiltonian, i.e. we consider a range of ansatz parameters $\boldsymbol{g}=\{m,\lambda\}$ to evaluate the constraints. As shown in \Fig{fig:Scheme1}, we obtain a landscape, where minimizing the absolute value $|\mathcal{C}|$ allows us to find the correct coupling parameters $ma =2$ and $\lambda a^2 = 0$, see appendix~\ref{sec:thermal-Gaussian}.

\begin{figure}[t!]
	\includegraphics[width=1.\columnwidth]{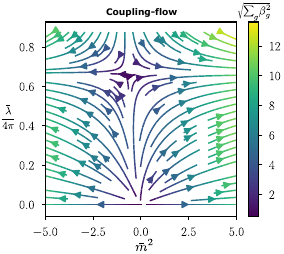}
 \caption{\textit{Resolution-scale flow of the learned Hamiltonian parameters (perturbative analytics).}  HL captures the flow of coupling parameters with the spatial measurement resolution as illustrated for the $\lambda\varphi^4$ theory in one dimension. Shown is the gradient of the dimensionless couplings $\bar{\lambda}=a^2\lambda/(2\pi)^2$, and $\bar{m}^2=a^2m^2/(2\pi)^2$ which determines the flow of coupling parameters with the resolution $a$. To compare couplings at different resolutions $a_1$ and $a_2$, we have $\lambda_1/\lambda_2 = a^2_2\bar{\lambda}(a_1)/[a^2_1\bar{\lambda}(a_2)]$ starting from an initial value $\lambda$ at the microscopic cutoff, $ \bar{\lambda}(a_0) = a_0^2 \lambda$. The absolute rate of change is given by the length of the vector $(\beta_{\bar{\lambda}}^2 + \beta_{\bar{m}^2}^2)^{-1/2}$, which becomes small in the vicinity of RG fixed points with~$\beta_{\bar{\lambda}} =\beta_{\bar{m}^2} = 0$.}
	\label{fig:RG}
\end{figure}

\subsection{Scale-dependence and RG flow in vacuum}
\label{sec:RG-flow}

In a quantum field theory, coupling parameters change with length scale, starting from an initial value at a cutoff scale $\Lambda$ and flowing towards their corresponding values at larger length scales, where they determine the properties relevant for macroscopic measurements. Similarly, the measurement resolution $a$ allows us to perform our Hamiltonian learning protocol at different spatial length scales to gain access to such a scale dependence, reminiscent of the renormalization group flow.

Here, we illustrate the observation of this flow in Hamiltonian learning, and for simplicity, we consider the continuum limit where the system Hamiltonian is given by a $\lambda\varphi^4$ scalar field theory with (hard) momentum cutoff $\Lambda$. We furthermore restrict ourselves to the mass $m$ and the interaction strength $\lambda$, while we neglect the generation of higher-order operators for simplicity. To generate the input data, we compute the correlation functions including the corrections they receive from quantum fluctuations perturbatively ($\lambda a^2_0 \ll 1$) up to order $\mathcal{O}(\lambda^2)$. These correlations are the basis for the evaluation of the constraint equations, which are minimized by the learned coupling parameters. As a consequence of the fluctuations of the inobservable modes $\eta$, which generate effective correlations between all other modes, the minima of the constraints shift depending on the resolution scale. Following this change allows us to extract the coupling flow of the theory via the (perturbative) shift of the learned couplings. As detailed in appendix~\ref{sec:RG}, we get the following shifts $\Delta g (a)=g(a)-g$ relative to the microscopic couplings $g$, 
\begin{align}
    \Delta m^2(a) &=  \frac{\lambda}{4\pi }\mathrm{log}\Bigg(\frac{\Lambda^2+m^2}{\left[\Lambda_a+\sqrt{\Lambda_a^2+m^2}\right]^2}\Bigg)\, , \\
    \Delta\lambda(a) &= -\frac{3\lambda^2 }{4\pi m^2}\left(\sqrt{\frac{\Lambda^2}{\Lambda^2+m^2}}-\sqrt{\frac{\Lambda_a^2}{\Lambda_a^2+m^2}}\right)\, ,
\end{align}
where $\Lambda_a = 2\pi/a$ is the energy scale associated with the measurement resolution. In accordance with the usual RG flow these shifts yield corresponding $\beta$-functions for the dimensionless coupling parameters, as $\beta_{g} = a\partial_a g$, which in this case read
\begin{align}
	\beta_{\bar{m}^2} &= 2 \bar{m}^2 + \frac{ \bar{\lambda}}{4\pi} \, ,\\
 \beta_{\bar{\lambda}} &= 2\bar{\lambda} - \frac{ 3\bar{\lambda}^2}{4\pi} \, ,
\end{align}
where we defined the dimensionless couplings as $\bar{\lambda}=a^2\lambda/(2\pi)^2$ and $\bar{m}^2=a^2m^2/(2\pi)^2$. The $\beta$-functions yield the flow shown in \Fig{fig:RG}. Motivated by the standard renormalization procedure, we considered here constraints on two-point functions and four-point functions respectively. While the mass shift is efficiently learned from the constraint on two-point functions with quadratic $\hat{\mathcal{O}}$, i.e. $f[\hat{\phi}] = \hat{\phi}_n$, the renormalization of the quartic coupling is best learned from the corresponding constraint with quartic $\hat{\mathcal{O}}$ and $f[\hat{\phi}] = \hat{\phi}_{n_1}\hat{\phi}_{n_2}\hat{\phi}_{n_3}$, see also appendix~\ref{app:choice-of-constraints}.

\subsection{High-temperature limit of thermal states}

In the limit of high temperatures, a quantum field theory reduces to a classical-statistical field theory, see appendix~\ref{sec:classical-limit} for details. The classical-statistical field theory will be the relevant scenario in our following analysis of the sine-Gordon model experimentally realized in tunnel-coupled superfluids. It also allows us to efficiently simulate the input data numerically for interacting theories with non-Gaussian correlations. As a consequence of the high-temperature limit several simplifications occur, which we will describe in this section.

For $\beta\rightarrow 0$ the path integral is dimensionally reduced to the expression
\begin{align}
Z^{(cl)}[J] = \int \mathcal{D}\varphi\, \exp(-\beta \mathcal{H}[\boldsymbol{g};\varphi] + J_n \phi_n) \, ,
\end{align}
where the fields in the Euclidean action are evaluated at the temporal zero modes, i.e. $\varphi_x = \int _0^\beta \mathrm{d}\tau\,\varphi(\tau,x)$, which yields the classical field theory Hamiltonian $\mathcal{H}[\boldsymbol{g};\varphi] = \int_x \big[  \frac{1}{2}(\nabla\varphi_x)^2 + V(\boldsymbol{g};\varphi_x)\big]$ that we want to learn. Equivalently, the dimensional reduction amounts to the decoupling of the momentum fields. In view of the application to the sine-Gordon theory in sections~\ref{sec:application} and \ref{sec:Experiment}, we have no mass term in the free part of the field theory, but consider such terms as part of the interaction $V$. This leads to the additional simplification that the optimization problem for the constraints remains linear even for the more realistic measurement models $W$ considered in these sections.

Similar to the quantum case before, we next integrate out the short-distance modes from the path integral resulting in an effective Hamiltonian~$\mathcal{H}[\boldsymbol{g};a;\phi]$. This Hamiltonian induces classical correlation constraints, which we compute using the Dyson-Schwinger equations, i.e. we extract the saddle point of the Hamiltonian via infinitesimal field transformations $\phi \rightarrow \phi + \epsilon$, and use that correlation functions of the saddle point with other field functionals $f$ average to zero, see appendix~\ref{sec:Schwinger-Dyson}. We obtain an infinite set of constraints for correlation functions, which reads $\mathcal{C}^{(cl)} = 0$, with
\begin{equation}
\label{eq:classical-constraint}
   \mathcal{C}^{(cl)}(\boldsymbol{g}) =  \left\langle f[\phi]\frac{\partial(\beta  \mathcal{H}[\boldsymbol{g};a;\phi])}{\partial \phi_n} - \frac{\partial f[\phi]}{\partial \phi_n}\right\rangle \, .
\end{equation}
We note that $\mathcal{C}^{(cl)}$ depends on  $n$ (and $f$), but we suppress this in the current notation where it is clear from the context.

\subsection{Summary of concepts}

The overall workflow of HL in QFTs is summarized in~\Fig{fig:Scheme1}. We distinguish three conceptual blocks as ingredients to our protocol: \emph{Quantum simulation} to experimentally generate correlation functions as input data; \emph{Hamiltonian flow} to formulate a suitable ansatz Hamiltonian at the measurement scale; and \emph{Classical post-processing}, where we relate the two to systematically constraint the set of Hamiltonian ansätze consistent with the data. At the same time, the learned couplings flow with the resolution scale and can therefore give more detailed insight into the flow of Hamiltonians with scale. 

In the following, motivated by the experimental realization of a sine-Gordon model with tunnel-coupled superfluids, we consider an application of this workflow to numerical data of a sine-Gordon field theory extracted from lattice simulations.

\section{Numerical demonstration for the sine-Gordon model}
\label{sec:application}

 In this section, we implement the workflow of \Fig{fig:Scheme1} numerically and extract the microscopic couplings of a one-dimensional classical-statistical field theory. To be specific, we consider again the sine-Gordon field theory with the corresponding Hamiltonian
 \begin{align}
 \label{eq:classical-Sine-Gordon}
     \beta\mathcal{H}_{sG} = \int_x \frac{\lambda_T}{4}\left[\frac{1}{2}(\nabla\varphi_x)^2 - \frac{1}{\ell_J^2}\cos(\varphi_x)\right]\, .
 \end{align}
The specific form of \eqref{eq:classical-Sine-Gordon} is motivated by the experimental realization of tunnel-coupled superfluids \cite{zache2020extracting}. The coupling constants correspond to length scales set by the thermal fluctuations ($\lambda_T$) as well as the tunnel-coupling of the superfluids ($\ell_J$) respectively. The effective interaction strength is set by the dimensionless ratio of these couplings, $Q=\lambda_T/\ell_J$, which can be tuned in the experiment via the tunnel-coupling as further explained below. 

 While in the previous discussion in section~\ref{sec:HL-theory}, we used a simplified model for the measurement $W$, we employ more realistic measurement models in the following. We stress that the choice of the measurement model affects the form of the effective Hamiltonian and hence the constraint equations.

\subsection{Method}
\label{sec:Numerical-demonstration}

We again generate the required input data for our protocol from an ``experimental" Hamiltonian~$\mathcal{H}_{\mathrm{sys}}$, which we simulate as a lattice discretized version of the Hamiltonian~\eqref{eq:classical-Sine-Gordon}. From the numerical simulations, we extract the corresponding thermal correlation functions.

Again, the lattice represents a UV-regularization scheme which introduces the lattice spacing $a_0$ as an explicit cutoff. In that sense, we simulate an experiment, which yields an effective sine-Gordon theory for length scales much larger than $a_0$. In the following, we will probe this field theory description on the resolution scale $a$ which we vary from small values ($a\sim \mathcal{O}(a_0)$) to values on the order of the correlation length (i.e. $a\sim \mathcal{O}(\ell_J)$).

In our simulations, we tune the interaction parameter $Q$ across a wide range of values, by fixing $\lambda_T \approx 17.35\mu\mathrm{m}$ and varying $\ell_J$ accordingly. 
To compute a sample of individual field configurations efficiently, we use a transfer matrix method detailed in Ref.~\cite{SchweiglerTransfer2018}. The thermal ensemble of the field theory is simulated on a microscopic (UV) lattice with 2,000 lattice points with spacing $a_{UV}=0.025\mu$m.
To compute the constraints $\mathcal{C}^{(cl)}$, we extract the relevant equal-time correlations $\langle \phi_{n_1} \cdots \phi_{n_m}\rangle$ of the coarse-grained field using an experimentally motivated model for $W$. 

Unless stated otherwise, numerical correlation functions are computed from averages over $N=$10,000 independent samples of field configurations. The learning results are shown as averages and standard deviations of 10 applications of our protocol to independent data sets.
 
 To assess the success of Hamiltonian learning, we consider a statistical $\chi^2$-test as a model-independent measure to observe the faithful realization of the correlation constraints $ \mathcal{C}^{(cl)}(\boldsymbol{g})=0$ without reference to the exact microscopic couplings (see appendix~\ref{sec:stat-analysis} for details). Values of $\chi^2/\nu \sim 1$ ($\nu$ counts the number of statistically independent degrees of freedom in the correlation constraints) confirm statistical agreement between the data and the field theory ansatz. Larger values signal that the data is not described by the ansatz. 

 To learn the Hamiltonian of the sine-Gordon theory, we consider the following class of ansatz Hamiltonians
\begin{align}
 \label{eq:classical-Sine-Gordon}
     \beta\mathcal{H}[\{\lambda_T,\boldsymbol{g}\},\varphi] = \int_x \frac{\lambda_T}{4}\left[\frac{1}{2}(\nabla\varphi_x)^2 + V(\boldsymbol{g};\varphi) \right]\, .
 \end{align}
 We identify the following tasks and provide an explicit numerical demonstration: First, our goal is to demonstrate the ability to learn the correct parameters by choosing the anticipated form of the ansatz Hamiltonian. Second, we will investigate the scale dependence of the learned parameters. Third, we want to differentiate between different ansatz Hamiltonians and explicitly identify $V(\boldsymbol{g};\varphi) \rightarrow -(1/\ell_J^2)\cos(\varphi_x)$.

\begin{figure}[t!]
	\includegraphics[width=.95\columnwidth]{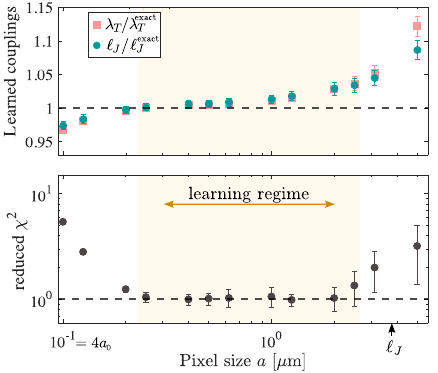}
	\caption{\textit{Scale dependence of Hamiltonian Learning of field theories (classical-statistical numerics).} (Top)~We learn the couplings $\lambda_T$ and $\ell_J$ of a relativistic sine-Gordon ansatz Hamiltonian from data generated by the sine-Gordon lattice Hamiltonian with exact couplings $\lambda_T^{\mathrm{exact}} = 17.35\mu$m, $\ell^\mathrm{exact}_J=2.70\mu$m. (Bottom)~Learning success is independently confirmed with reduced $\chi^2$-values of order one, testing the fulfillment of the correlation constraints within statistical fluctuations for the considered sample size. Here, we average over $10$ applications of our protocol, each based on correlation functions from $10,000$ independent samples (error bars represent standard deviation). 
 }
	\label{fig:Scheme2}
\end{figure}

\begin{figure*}[t!]
	\includegraphics[width=1.95\columnwidth]{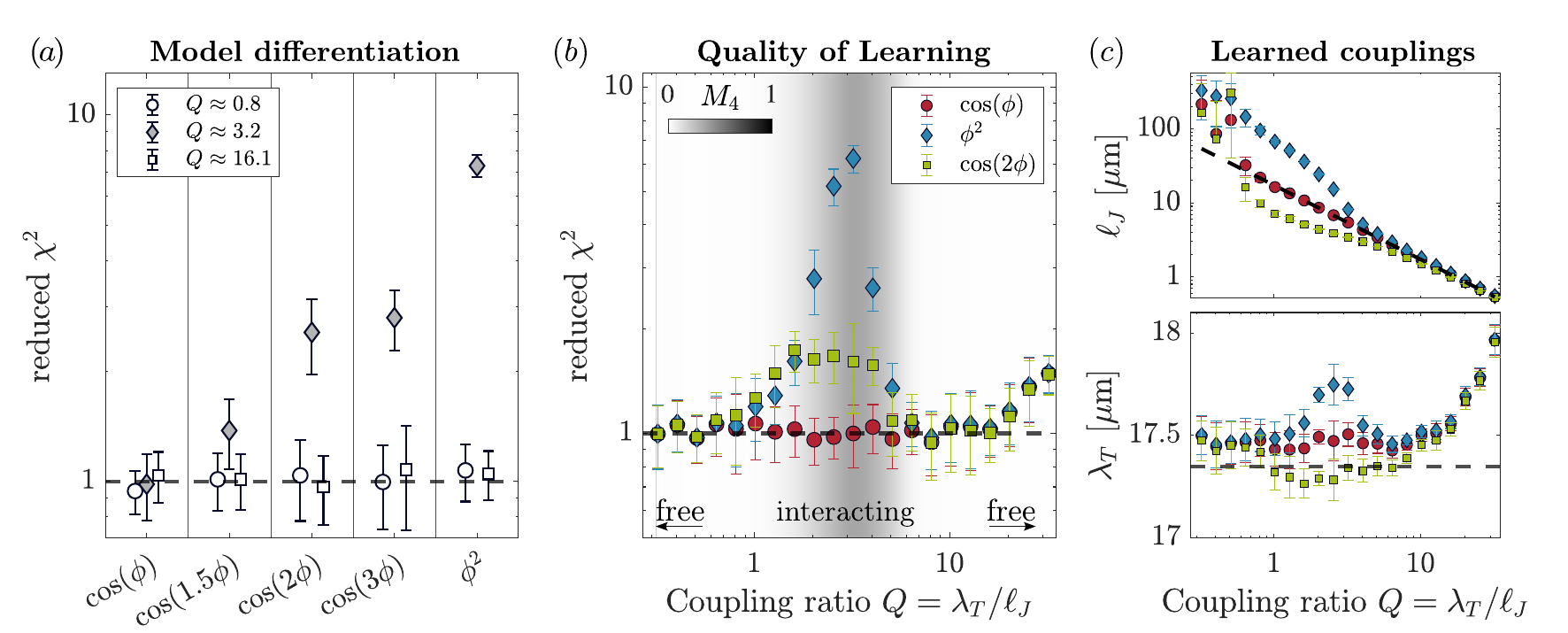}
	\caption{\textit{Learning field theory potential (classical-statistical numerics).} ($a$)~The reduced $\chi^2$ signals the agreement of data and field theory ansatz for three regimes of the coupling ratio~$Q=\lambda_T/\ell_J$. For each ansatz $V$, we choose $f\propto\partial V/\partial \phi$. The protocol allows us to clearly distinguish different potentials at intermediate values of $Q$, while all potentials agree on the same free field theory in the two limits of large and small $Q$ as expected. ($b$)~We learn the optimal coupling parameters for three ansätze across a range of coupling ratios $Q$, where we keep $\lambda_T \approx 17.36 \mu$m fixed and tune~$\ell_J$. The dark-shaded region signals the interacting regime with $M_4>0$, where we expect the cosine potential to be relevant (see main text). Here, we consider a fixed pixel size $a=0.5\mu$m. For $Q \lesssim \lambda_T/a \approx 34.7$ (i.e. $\ell_J > a$), the relevant physics is resolved and the microscopic couplings can be reconstructed. ($c$)~Learned couplings approximately agree with exact values for regions with reduced $\chi^2 \approx 1$. We use the same constraint $f = [\sin{\phi} + \phi + \sin(2\phi)]$ for all data in ($b$) and ($c$).
 }
	\label{fig:Learning}
\end{figure*}
\subsection{Scale-dependence of sine-Gordon couplings}

The successful learning of the microscopic couplings from numerical data with our protocol is first illustrated in \Fig{fig:Scheme2}. To illustrate Hamiltonian learning for different resolution scales in this case, we first focus on the ansatz Hamiltonian with the correct functional form 
\begin{align}
V(\ell_J;\varphi) = -\frac{1}{\ell_J^2}\cos(\varphi_x)\, .
\end{align}
The learned coupling parameters for different resolution scales $a$ are shown in the top panel. Specifically, the learned couplings are compared to the exact coupling parameters of the lattice simulation. 

To constrain the coupling parameters, we choose the constraint $\mathcal{C}^{(cl)}$ with $f[\phi] =\sin(\phi_m)$, in agreement with the symmetries $\phi \rightarrow \phi + 2\pi$, and $\phi \rightarrow -\phi$ of the ansatz. The results of our $\chi^2$ analysis are shown in the bottom panel: our protocol successfully learns the couplings for an intermediate ``learning regime'', where $a_{0}\ll a \ll \ell_J, \lambda_T$.

For $a\sim a_{0}$, we see deviations due to the UV-cutoff in our data-generating model, i.e., at that scale, the data does not represent a system in the continuum limit. Here, this originates from discretization effects in our numerical simulations.
Conversely, at scales $a \sim \ell_J, \lambda_T$, the value of the reduced $\chi^2$ signals the end of the learning regime. To learn the Hamiltonian in this regime, we would have to update its operator content, as the effective Hamiltonian receives (possibly non-local) corrections beyond those considered in the present ansatz. In general, finding these further corrections is non-trivial, such that in general Hamiltonian learning may become inefficient in this case.

Even within the learning regime, that is reduced $\chi^2\sim 1$,  we observe small changes in the coupling parameters with respect to the exact input parameters. The deviations can be understood as fluctuation effects that generate effective and scale-dependent, interactions (see also ref.~\cite{zache2020extracting}). Within the learning regime, this leads to a renormalization of the coupling parameters, similar to the coupling flows discussed in section~\ref{sec:RG-flow}.

\subsection{Model differentiation}
In addition to learning the couplings of our model, Hamiltonian learning allows us also to distinguish between different potentials in the ansatz Hamiltonian. In \Fig{fig:Learning}, we give the learning results for various choices of the potential $V$ across different interaction ranges of the sine-Gordon data. Panel ($a$) demonstrates that different potentials are clearly distinguished for $Q = \lambda_T/\ell_J \approx 3.2$, where the sine-Gordon potential $V\sim\cos(\varphi)$ is favoured over other choices such as $\cos(1.5\varphi)$, $\cos(2\varphi)$, $\cos(3\varphi)$, and~$\varphi^2$. To distinguish these potentials, we use different constraints, $f \sim \partial_{\phi_{m}} V(\phi_m)$. The linear constraint $f=\phi_m$ is very sensitive to phase windings in the sine-Gordon theory, and therefore the quadratic theory can be faithfully excluded. 

Conversely, we find that the data is represented equally well by all potentials in the cases $Q\approx 0.8$, and $Q\approx 16.1$. Indeed, in both these limits the sine-Gordon theory is well-approximated by a free field theory. This is confirmed by the quantity $M_4$~\cite{schweigler2017experimental}, which quantifies the mean size of the connected (subscript $\textbf{c}$) four-point correlation function i.e. $M_4 = \sum_n \langle (\phi_n - \phi_{n_0})^4\rangle_\textbf{c} / \sum_n\langle (\phi_n - \phi_{n_0})^4\rangle$, where $n_0$ is an arbitrarily chosen reference point. For Gaussian thermal states, corresponding to free field theories, this measure reduces to zero.

The coupling dependence of the learned field theory – along with the corresponding values of~$M_4$ – is illustrated in \Fig{fig:Learning}($b$), where we show a detailed scan over the coupling ratio~$Q$. The interacting regime (intermediate $Q$, dark shading representing non-zero $M_4$) is clearly identified as the coupling regime which is in statistical disagreement with the quadratic theory~$V\sim \varphi^2$. Similarly, other potentials fail (e.g. $\cos(2\varphi)$), while the sine-Gordon potential yields good agreement overall within the ``learning regime'', i.e. for $ a\lesssim \ell_J$, corresponding to $Q\lesssim \lambda_T/a$. Panel ($c$) confirms the learning of the correct coupling values in this regime, up to a shift at the percentage level. This deviation agrees with the results of \Fig{fig:Scheme2}($b$), and originates from the imperfect effective theory ansatz at a finite resolution scale and it may be interpreted as a perturbative renormalization of the couplings.

\begin{figure}
	\includegraphics[width=1.\columnwidth]{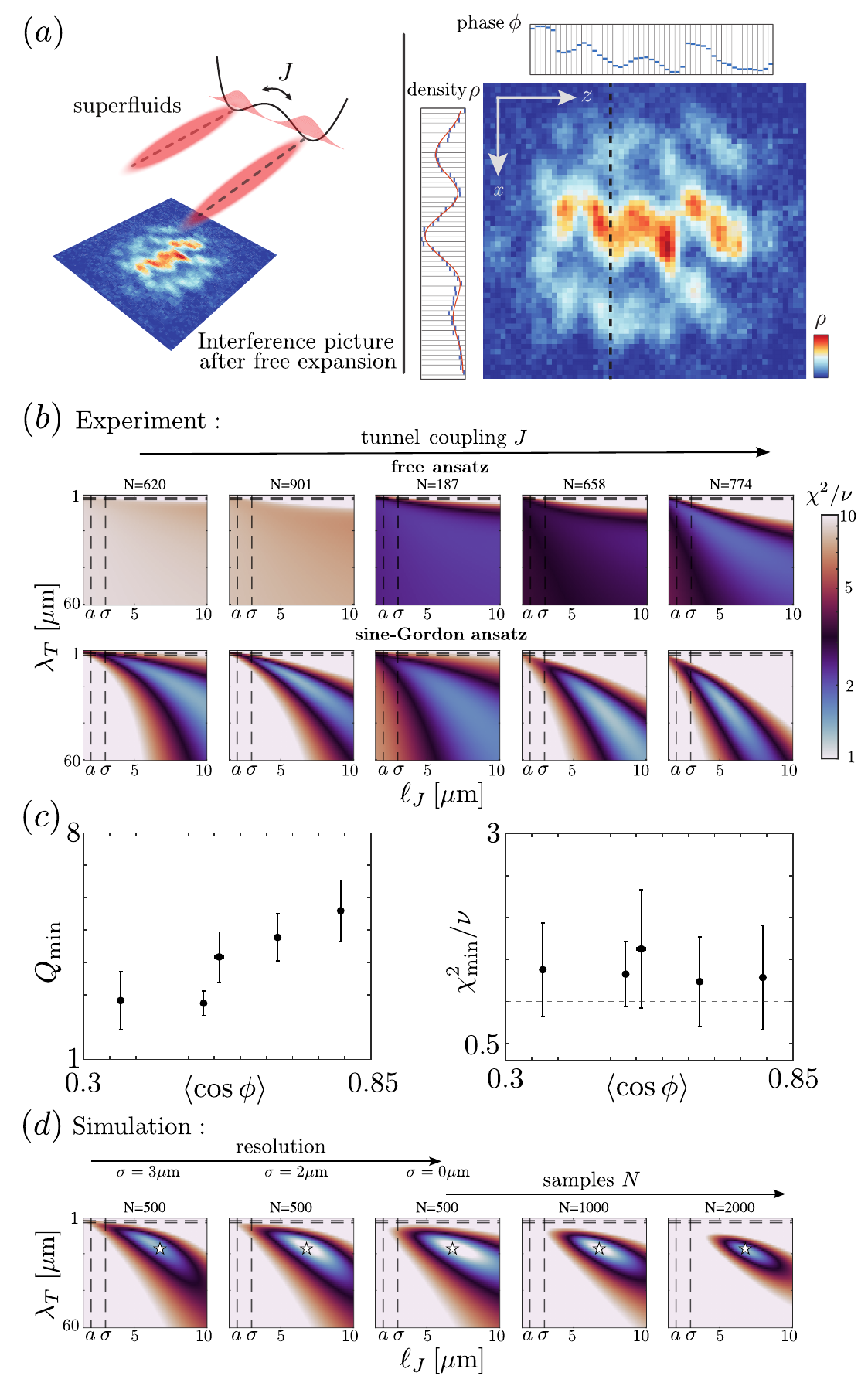}
	\caption{\textit{Learning the sine-Gordon Hamiltonian for tunnel-coupled superfluids (experimental data).}~($a$) The experimental setup consists of two tunnel-coupled superfluids on an atom chip which are interfered and imaged onto a CCD camera. From the displacement of the interference density ($\rho$) maxima along pixel $i$ (dashed line) we extract the coarse-grained phase profiles $\phi_i$ for pixels $i = 1,...,N_L$ and $N$ realizations of the experiment. The measurement introduces additional scales through imaging resolution $\sigma = 3\mu$m and pixel size $a=2\mu$m. ($b$) We probe the correlations for several data sets with different ansatz Hamiltonians, a free theory and a sine-Gordon theory. For each data set and ansatz, we show the reduced $\chi^2$ of the Hamiltonian constraint with $f[\phi] \propto \partial V(\phi)/\partial \phi$. Within the given accuracy of the measurements, the data is consistent with the couplings in violet areas, thus the sine-Gordon field theory is favored over the quadratic potential. ($c$) For the sine-Gordon ansatz, the minimal values for $\chi^2/\nu$ and learned coupling ratio $Q$ are shown below. Statistical error bars are estimated from a jackknife resampling of 150 samples respectively. ($d$) Sine-Gordon Hamiltonian learning for simulated data with exact couplings $\lambda_T\approx 17.35\mu$m and $\ell_J\approx 6.78\mu$m (white star). Shown is the effect of improving the resolution from $\sigma = 3\mu$m to $\sigma = 0$ (at fixed $a=2\mu$m), which yields an increased region of $\chi^2/\nu\approx 1$. Subsequently, the number of samples is increased from $N=500$ to $N=2,000$, resulting in improved restrictions for the optimal coupling parameters.}
	\label{fig:Experiment}
\end{figure}

\section{Hamiltonian Learning for a quantum gas experiment}
\label{sec:Experiment}
We now apply our Hamiltonian learning protocol to an experiment of a cold atomic gas realizing a field theory in the continuum. In particular, we demonstrate the ability to learn the effective Hamiltonian from experimental data from thermal states of two quasi--one-dimensional tunnel-coupled superfluids of $^{87}$Rb atoms in a magnetic double-well potential at $20\,$nK, see \Fig{fig:Experiment}($a$). The experimental setup is described in detail in Ref.~\cite{schweigler2017experimental}.

In such a setup, on the microscopic level, the atoms are represented by bosonic operators $\hat{\psi}^{\dagger}_i(x)$ and $\hat{\psi}_i(x)$ for each well $i=1,2$. In general, each field operator involves a radial density, as well as an angular phase degree of freedom. However, in the experimentally realized parameter regime of density and temperature, the dynamics is reduced to a classical-statistical description, where density and phase decouple, and all non-trivial statistics of the fields is limited to the relative phase field $\varphi(x)$. The tunneling term between the two wells with strength $J$ gives rise to a periodic cosine potential, $\sim J\cos\varphi(x)$~\cite{schweigler2017experimental}. Hence, we expect the data to be described by a sine-Gordon field theory~\cite{gritsev2007linear}.

This relative phase is experimentally extracted via matterwave interference from time-of-flight images, see \Fig{fig:Experiment}($a$). Here, we focus on a central region of length $L\approx 50 \mu$m to minimize the inhomogeneity of the trapping potential. The measurement process introduces two scales: the finite imaging resolution, modeled by a Gaussian convolution with $\sigma = 3\,\mu$m, and the pixel size $a=2\mu$m of the CCD camera – both of which we take into account in our definition of~$W$, see the appendix~\ref{sec:measurement-model}. 
These scales set the relevant length scales on which we aim to test the emergence of a field theory Hamiltonian.

In \Fig{fig:Experiment} we show the results of our learning scheme applied to data from several different parameter sets within the interacting regime of intermediate $Q$. We compare the Hamiltonian protocol for two different ansätze in panel ($b$): the free field theory (top) and the sine-Gordon field theory (bottom). We compare both cases using different constraint functionals $f$ corresponding to the different symmetries of the theories. For the sine-Gordon theory, the field $\phi$ represents an angular variable on a compact interval $[0,2\pi]$, and hence we choose a periodic constraint $f[\phi] = \sin(\phi)$. In the case of the free field theory, however, the fields live on non-compact intervals, and we use constraints involving non-periodic correlation functions, e.g. constraints corresponding to $f[\phi] = \phi$. They are particularly sensitive to the phase windings of $\phi$. To construct the constraint, we also employ a spatial convolution procedure designed to mitigate finite-size effects, as detailed in the appendix~\ref{sec:measurement-model}.

The results are shown in \Fig{fig:Experiment}($b$) where we show the reduced $\chi^2$ values for different data sets and a range of coupling parameters $\lambda_T$ and $\ell_J$ for both ansätze. For the free field theory, the reduced $\chi^2$ values significantly deviate from one for all ansatz parameters. Conversely, for the sine-Gordon ansatz, we find coupling parameters leading to $\chi^2/\nu \sim 1$ for all data sets. The optimal coupling parameters, along with the corresponding minimal $\chi^2$-values $\chi^2_\mathrm{min}/\nu$, are displayed in \Fig{fig:Experiment}($c$) [in the order of the data sets in ($b$)], and we find $Q$-ratios in the range of $1\lesssim Q \lesssim 6$, corresponding to the interacting regime (see also~\Fig{fig:Learning}). Specifically, we find the expected monotonous growth of the coupling ratio $Q$ with the mean value $\langle\cos(\phi)\rangle$~\cite{schweigler2017experimental}.

The Hamiltonian learning performance could be further improved with higher spatial resolution as well as a larger sample size. This is illustrated in \Fig{fig:Experiment}($d$), where we show numerical simulations of the protocol with realistic experimental parameters. Using the same constraint, we test the influences of both improvements in the resolution $\sigma$ (from $3\mu$m to $0\mu$m while keeping $a=2\mu$m fixed) and the sample size $N$ (from $N=500$ to $N=$2,000). In combination, these improvements lead to a better estimate of the exact coupling parameters of the input data (white star). The statistical sample size plays an integral role in our Hamiltonian learning scheme. We thus investigate its influence on learning performance in more detail in the next section.

\section{Statistical sample complexity}
\label{sec:statistics}

\begin{figure}[t!]
	\includegraphics[width=1.02\columnwidth]{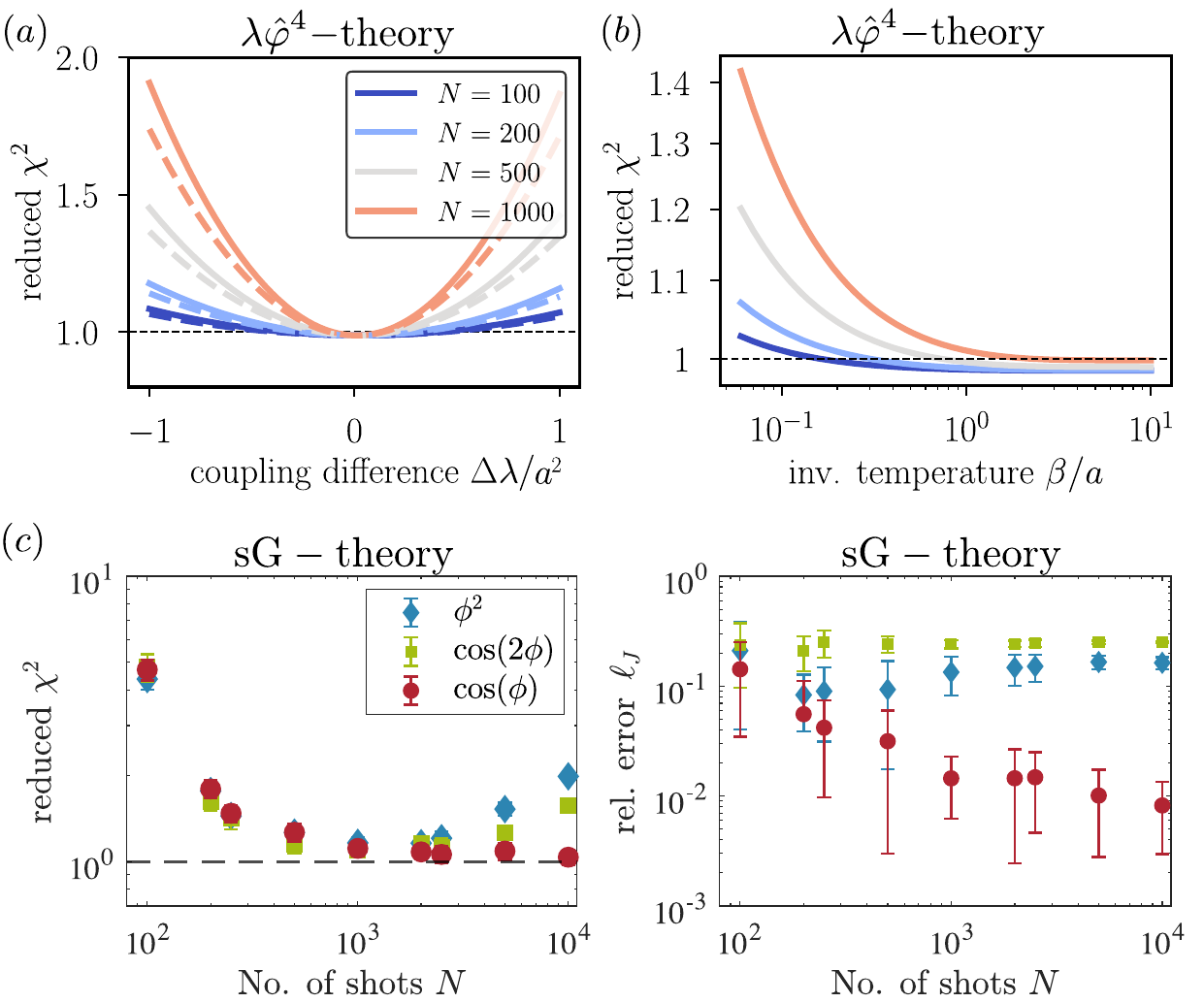}
	\caption{\textit{Statistical dependence of Hamiltonian learning for two cases: exact analytics and classical-statistical numerics.} (Top panels) Analytic calculation for the case of Gaussian field correlations with an interacting (perturbative) ansatz [with "true" mass parameter $ma=Ma = \pi$]. ($a$) For the "true" ansatz, with exact scale-dependant Hamiltonian in this case, the reduced $\chi^2$ stays at 1 (for finite statistics there is a small offset), otherwise, it diverges with increasing measurement precision and inverse temperature (solid line: $\beta/a = 1$, dashed line: $\beta/a = 1.5$). ($b$) At a fixed coupling $\lambda a^2 = 0.2$, we compute the reduced $\chi^2$ for various temperatures indicating that the learning procedure is more restrictive at large temperatures in this case. (Bottom panels) Classical-statistical simulation of the sine-Gordon theory with parameters as in \Fig{fig:Scheme2}($c$) We show the reduced $\chi^2$ for three different ansatz Hamiltonians with increasing statistics in the classical statistical case. Here, we average data points over $10$ instances of applying our protocol to sets of $N$ realizations of field configurations; error bars are standard deviations. We find agreement for all three potentials for intermediate numbers of realizations of order $\sim 10^3$. Increasing the statistics allows to distinguish between different potentials. The relative error of the coupling constant reaches a plateau for incompatible ansatz Hamiltonians. For potentials with good $\chi^2$-agreement, the relative error keeps decreasing. Here, we choose constraints~$f\propto V'$.}
	\label{fig:Statistics}
\end{figure}

We first consider the simple case of Gaussian input data, corresponding to the correlation functions of a free, thermal quantum field theory. The dependence of the reduced $\chi^2$ on the number of measurements is shown in \Fig{fig:Statistics} for a range of coupling parameters $\lambda$ in the ansatz of a $\phi^4$ scalar field theory. In the present case, the modes decouple and we compute the expected $\chi^2/\nu$-values analytically (see appendix~\ref{sec:thermal-Gaussian}). Since the effective, scale-dependant Hamiltonian is quadratic, one achieves $\chi^2/\nu = 1$ exactly at the correct microscopic parameters. Away from these optimal parameters, $\chi^2$ grows with sample size and temperature as expected. Notably, our results indicate that the performance of Hamiltonian learning improves at higher temperatures. This is different to the case of discrete spin systems, where the learning performance reduces with temperature~\cite{anshu_sample-efficient_2021}. In general, we expect the dependence with temperature to depend on the details of the interaction potential. In the case of bosonic modes with infinite-dimensional Hilbert spaces, the infinite temperature limit is ill-defined in the sense that typical observables diverge as $T\rightarrow \infty$; higher temperatures allow to probe the potential at larger field values, hence yielding more information on the coupling parameters. At a certain temperature, however, the identification with the field theory breaks down and the task of Hamiltonian Learning reduces to the microscopic description of the quantum simulator.

In order to investigate the more general case of interacting, non-Gaussian correlations, we again consider the classical-statistical theory, where different modes do not decouple. In \Fig{fig:Statistics}($c$) we demonstrate the performance of the Hamiltonian learning protocol for three different ansatz potentials and for increasing statistical sample size; there is no additional dependence since the temperature drops out in the classical-statistical limit. For $N\approx 1$,000 statistical samples, we observe successful learning, and all three field theories are consistent with the correlation data. By including more samples, we are able to observe that the potentials $V\sim \phi^2, \cos{2\phi}$ are inconsistent with the data, while the sine-Gordon field theory continues to be a good description of the physics. The inset shows the relative error of the learned coupling $(\ell_J-\ell^\mathrm{exact}_J)/\ell^\mathrm{exact}_J$: while the error for the other potentials saturates at a given sample size, it keeps decreasing for the ansatz $\cos{\phi}$ for all considered sample sizes, up to~$N=10$,000.
 From these considerations, the error for the ansatz $\cos{\phi}$ is expected to reach a plateau for even larger sample sizes, and the value of $\chi^2/\nu$ is expected to rise above $1$, correspondingly.

  While the absolute sample size to be considered depends on microscopic details, the behavior found in \Fig{fig:Statistics}($c$) is expected to also hold in general, e.g. for the experimental data considered in section~\ref{sec:Experiment}: the learned field theories are effective descriptions of the underlying physics, and hence they are not required to hold indefinitely – but increasingly high statistics are necessary to resolve the microscopic corrections of the physics at ever smaller length scales. In that sense, a finite statistical sample size is a fundamental ingredient to our identification of field theories in quantum simulators: increasing it reduces the number of field theory models consistent with the data. In this spirit, a verification of field theory quantum simulations via Hamiltonian learning is only sensible with respect to the given statistical sample size. Field theories should be learned with a sample size similar to the data used for the evaluation of physical questions in quantum simulations.

\section{Conclusion}
\label{sec:conclusion}

In summary, we have presented a protocol to learn quantum and classical Hamiltonian field theories from the data obtained from continuous quantum systems in both equilibrium and non-equilibrium settings.
In contrast to simple fitting of individual coupling parameters of a known Hamiltonian, Hamiltonian learning employs many-body correlations to systematically reveal the operator content of the underlying field theory Hamiltonian.
Using both analytical calculations and numerical simulations we explored the emergence of field theories and the flow of their coupling constants versus measurement resolution. By applying the protocol to experimental data of tunnel-coupled superfluids we learned the Hamiltonian of a relativistic classical field theory in thermal equilibrium.

Our experimental findings are consistent with the emergence of a sine-Gordon field theory. While the results show the qualitatively expected features, they further indicate the experimental challenges with respect to the required system size, measurement resolution and sample size. Combined with improved measurements, our protocol will allow us to quantify the agreement of the experimental data with the conjectured sine-Gordon theory more precisely and will give rise to measuring Hamiltonian flows in the experiment. To learn the Hamiltonian in more general settings, i.e. without explicit reference to specific quantum states, the experimental setting can be extended to perform quenches and therefore access the real-time dynamics of the effective field theory. Such an extension could allow us to investigate the emergence of effective field theories far away from thermal equilibrium~\cite{calzetta2009nonequilibrium}, for instance, to investigate dynamical critical phenomena, or to see the emergence of different effective Hamiltonians with time.

In general, to make full use of Hamiltonian learning in the quantum regime requires measuring quantum correlations between both fields and canonical momenta. Such correlations are not straightforwardly accessible in experiments but can be obtained via generalized measurements based on positive operator-valued measures (POVMs)~\cite{kunkel2019simultaneous}. These schemes allow to independently access both conjugate fields with spatial resolution, but will be limited by the vacuum fluctuations of an ancilla state. For learning the quantum Hamiltonian close to the vacuum state the required correlations can be inferred by tomography employing unitary rotations between fields~(see Refs.~\cite{gluza2020quantum,tajik2023verification} for the example of second-order correlations). For general, higher-order field correlations independent local rotations between the field quadratures are required (cf. Ref.~\cite{cooper2022engineering} for a recent demonstration). In this context, our work also relates to recently developed randomized measurements techniques~\cite{elben2023randomized}, which could similarly be applied to quantum field theory, and facilitate simultaneous evaluation of the many required correlation functions~\cite{huang2020predicting}. Such measurements would also give rise to further extensions of the methods presented here, to include incoherent interactions~\cite{pastori2022characterization,bairey2020learning} or to study the entanglement structure of quantum field theories~\cite{kokail2021entanglement,joshi2023exploring,calabrese2004entanglement}.

Besides these applications, our methods extend to other degrees of freedom ranging from multi-component bosons, spins, and fermions to emergent gauge fields or even anyons in higher-dimensional setups, as relevant for recent experiments~\cite{prufer2020experimental,frolian2022realizing,semeghini2021probing}. 
In two or higher dimensions, field theory ansätze are typically restricted even further: interaction terms are classified in terms of relevant and irrelevant operators, such that for sufficient separation of scales only a few local interactions remain as candidates. Our scheme provides a method of extracting these relevant operators and opens up new ways of addressing their resolution-scale flows directly from experimental data.

\section{Acknowledgments}
We thank T.\,Schweigler for experimental assistance taking some of the original data used in the experimental section. We thank J.\,Berges for discussions and collaboration on related work.
This work is supported by the European Union's Horizon Europe research and innovation program under Grant Agreement No. 101113690 (PASQuanS2.1), the ERC Starting grant QARA (Grant No.~101041435), the ERC-AdG {\em Emergence in Quantum Physics} (EmQ) under Grant Agreement No. 101097858 and the DFG/FWF CRC 1225 'ISOQUANT', (Austrian Science Fund (FWF) I~4863). M.P. has received funding from the European Union’s Horizon 2020 research and innovation program under the Marie Skłodowska-Curie grant agreement No 101032523 and from Austrian Science Fund (FWF): ESP~396 (QuOntM).

\clearpage
\appendix

\section{Effective action}
\label{sec:scale-dependent-quantum}
In this section, we expand on the details of integrating out the short-distance degrees of freedom in the quantum theory. The correlations of quantum fields can be accessed through the path integral
\begin{equation}
    Z = \int \mathcal{D}\varphi \, e^{-S}
\end{equation}
where $S$ is the Euclidean action. In our case, we consider a $\lambda \phi^4$ theory, and we get
\begin{align}
S = \int \mathrm{d}^2x \big[\frac{1}{2} (\partial \varphi)^2 - \frac{m^2}{2} \varphi^2 + \frac{\lambda}{4!} \varphi^4\big] \, ,
\end{align}
with partial derivatives $\partial_\mu$ with respect to coordinates $\mu = 0,1$ for temporal and spatial directions.

In our example, we choose an "imaging" kernel 
\begin{align}
    W_{nx} = \frac{1}{\sigma}\mathrm{sinc}[(x-na)/\sigma] ,
\end{align}
where we take $a=\sigma$ for simplicity. In Fourier space, the kernel reduces to a hard cutoff of Fourier modes at $p_{cut}= 2\pi/\sigma$.

Correlation functions can be accessed from derivatives acting on $Z[J]$, and subsequently evaluating the result at $J=0$. We introduce the sources by shifting $S\rightarrow S + \sum_{n}J_n \int_x W_{nx}\phi_x$, where the coupling of fields to the sources $J_n$ via the matrix $W$ encodes the imaging being limited to coarse-grained modes.

Using this, our field space separates into modes $\phi$ with support on momenta $p<2\pi/\sigma$ and $\eta$ with support on momenta $p>2\pi/\sigma$. As a consequence, the action splits as
\begin{align}
    S[\phi] = S[\phi] + S[\eta] + S_\mathrm{int}[\phi,\eta]\, .
\end{align}
At leading order, the fields decouple, and we get the action
\begin{align}
    S = \int \mathrm{d}\tau \sum_{n} \Big[\frac{1}{2}\dot{\phi}^2_n + \frac{1}{2}\phi_m M_{mn} \phi_n  + \frac{\lambda}{4!}\phi^4_n \Big] \, ,
\end{align}
where the "mass" matrix is given by
\begin{align}
    M_{mn}&=\int_{0}^{\frac{2\pi}{a}} \frac{\mathrm{d}p}{2\pi} (m^2 + p^2) e^{ipa(n-m)} \nonumber\\
    &= \left(\frac{\pi^2}{a^3}+ \frac{m^2}{a} \right)\delta_{nm} +\frac{2}{a^3}\frac{(-1)^{n-m}}{(n-m)^2} \, .
\end{align}
Here, $p^2+m^2$ is the typical energy-momentum relation, which we obtain from the Fourier transform of the derivative operator $G_0^{-1}=-\partial^2+m^2$ from the quadratic part of the microscopic action.
The action can be quantized to yield the Hamiltonian~(\ref{eq:ansatz}) stated in the main text.

\section{Choice of constraints and systematic learning of interactions}
\label{app:choice-of-constraints}
To choose the constraints, as often in field theory, we consider the symmetries of the measured degrees of freedom as well as the ansatz potentials. In the general case, we first prescribe a symmetry, for example a $\mathbb{Z}_2$ symmetry $\varphi\rightarrow -\varphi$ for the case of a $\lambda \varphi^4$ theory, or a field periodicity $\varphi \rightarrow \varphi +2\pi$ for the sine-Gordon theory. Subsequently, both the field potentials $V(\varphi)$ as well as the field functionals $f[\varphi]$ are chosen according to that symmetry. 

As an example for the detection of explicit symmetry-breaking interaction terms, we illustrate the learning of such couplings from correlation functions of a $\lambda\varphi^4$ theory with the corresponding $\mathbb{Z}_2$ symmetry $\varphi\rightarrow -\varphi$. As an example, we illustrate that HL yields a zero coupling for symmetry breaking interactions of the form $V_\mathrm{sb}=\lambda_3\varphi^3$ in the ansatz. The symmetry implies that all correlation functions of an odd number of fields vanish, i.e. $\langle \phi^{(2n+1)}\rangle =0$ for $n\in \mathbb{N}_0$. This implies, that for an even choice of $f$, $f[-\phi] = f[\phi]$ all symmetric interactions $\sim m^2\phi^2, \lambda\phi^4, ...$ drop out from the constraint. What remains is given by $\mathcal{C} = \langle f[\phi]\partial_{\phi_{n}}V^{\mathrm{eff}}_\mathrm{sb}\rangle\propto \lambda_3\langle f[\phi]\phi_n^2\rangle $. The remaining correlation function is even in $\phi$ and therefore in general non-zero. Consequently, the only possibility to satisfy this constraint ($\mathcal{C}=0$) is by setting $\lambda_3 = 0$. 

Under the principles of effective field theories, we would like to start from the most general Hamiltonian which is compatible with these symmetries. We may then optimize the ansatz by systematically constraining the coupling values. To this end, we propose to choose multiple constraints involving various field functionals $f$.

We illustrate the learning of these different interactions from the corresponding various constraints in the case of a weakly interacting scalar field theory with quartic ($\lambda\phi^4$) coupling. To learn the Hamiltonian, we first start from the simplest ansatz: a free theory with ansatz mass $m$ as discussed in the previous subsection. Using the quadratic constraint, $\sum_{nm}\langle\hat{\pi}_{n}\hat{\pi}_{m} + a\sum_{l} M_{ml} \hat{\phi}_{n}\hat{\phi}_{l}\rangle =0$, the leading order effects of the interaction $\lambda$ can be absorbed into the mass and therefore will lead to a slight shift compared to the original mass in the Hamiltonian. However, once we consider the quartic constraint with $f\sim\mathcal{O}(\hat{\phi}^3)$, i.e. $\mathcal{C} = \langle\hat{\phi}_{n_1}\hat{\phi}_{n_2}\hat{\pi}_{n_3}\hat{\pi}_{m}\rangle_{\mathrm{sym}} + a\sum_{l} M_{ml} \langle\hat{\phi}_{n_1}\hat{\phi}_{n_2}\hat{\phi}_{n_3}\hat{\phi}_{l}\rangle $, our set of constraint cannot be satisfied with the quadratic ansatz any more. That is, because the four-point functions contain contributions of connected four-point functions, e.g. $\langle\hat{\phi}_{n_1}\hat{\phi}_{n_2}\hat{\phi}_{n_3}\hat{\phi}_{l}\rangle_\textbf{c}$, at order $\mathcal{O}(\lambda)$. While the disconnected part of the constraint reduces to the previous two-point constraint, which fixes the learned mass and therefore $M$, this new contribution is not canceled without extra terms in the constraint. Therefore, we extend the ansatz to include additional interaction terms, starting with the $\lambda\hat{\phi}^4$ interaction in this case. This introduces a new term $\sim \lambda\langle f(\hat{\phi}) \hat{\phi}_m^3\rangle$ to the constraint, which cancels the remaining terms of the connected four-point functions to order $\mathcal{O}(\lambda^2)$. 

This approach can be extended step-by-step to systematically expand the Hamiltonian to include more and more interaction terms. In general, the learned couplings will get contributions from both the ``correct" value, as well as further shifts from all interactions at higher order, similar to the mass shift discussed here. These shifts depend on the resolution scale, on which we learn the Hamiltonian, and are therefore reminiscent of the renormalization group flow, see section~\ref{sec:RG}.

\section{Renormalization group flow from HL}
\label{sec:RG}
In this section, we describe the flow of coupling constants with the resolution scale, and how it can be obtained with Hamiltonian learning. At first, we start with the analytical expressions for the field operators, and correlation functions in the limits of free quantum fields as well as a perturbative $\lambda \varphi^4$ theory.

\subsection{Constraints for free fields}
We begin with the free-field expansion
\begin{align}
    \varphi(x) &= \int\frac{\mathrm{d}p}{2\pi} \frac{1}{\sqrt{2\omega_p}} (e^{ipx} a_p + e^{-ipx} a_p^\dagger)\, ,\\
    \pi(x) &= \int\frac{\mathrm{d}p}{2\pi} (-i)\sqrt{\frac{\omega_p}{2}} (e^{ipx} a_p - e^{-ipx} a_p^\dagger)\, ,
\end{align}
where $[a_p,a_q^\dagger] = (2\pi)\delta(p-q)$ and including the single-particle energy $\omega_p = \sqrt{p^2+m^2}$. The time-dependent Heisenberg field operators are given by
\begin{align}
    \varphi(x,t) &= \int\frac{\mathrm{d}p}{2\pi}\frac{1}{\sqrt{2\omega_p}} (e^{ipx-i\omega_p t} a_p + e^{-ipx+i\omega_p t} a_p^\dagger)\, , \\
    \pi(x,t) &= \int\frac{\mathrm{d}p}{2\pi}(-i) \sqrt{\frac{\omega_p}{2}}(e^{ipx-i\omega_p t} a_p - e^{-ipx+i\omega_p t} a_p^\dagger)  .
\end{align}

In the free vacuum theory, one obtains the following correlation functions
\begin{align}
    \langle \varphi(x,t)\varphi(y,t') \rangle &= \int \frac{\mathrm{d}p}{2\pi}\frac{1}{2\omega_p}  e^{ip(x-y)-i\omega_p(t-t')}\, ,\\ 
    \langle \pi(x,t)\pi(y,t') \rangle &=- \int \frac{\mathrm{d}p}{2\pi}\frac{\omega_p}{2}  e^{ip(x-y)-i\omega_p(t-t')}\, ,\\ 
    \langle \varphi(x,t)\pi(y,t') \rangle &= \frac{1}{2}\int \frac{\mathrm{d}p}{2\pi}  e^{ip(x-y)-i\omega_p(t-t')}\, .
\end{align}

As an example, we first consider the quadratic constraint, which is given by the Schwinger-Dyson equations
\begin{align}
    \langle\mathrm{T} [G^{-1}\varphi(x,t)]\varphi(y,t')\rangle \, ,
\end{align}
which is equivalent to the constraint 
\begin{align}
    G^{-1}\langle\mathrm{T} \varphi(x,t)\varphi(y,t')\rangle &= -i\delta(x-y)\delta(t-t')\, ,
\end{align}
and where $G^{-1} = -\partial^2 - m^2 $. From this, we obtain at equal times
\begin{align}
   - \langle \pi(x,t)\pi(y,t)\rangle - (\nabla^2_x + m^2) \langle \varphi(x,t)\varphi(y,t) \rangle = 0.
\end{align}
This is indeed true for the free theory, which we can confirm by plugging in the above correlation functions
\begin{align}
 \int \frac{\mathrm{d}p}{2\pi}\frac{\omega^2_p - (p^2+m^2)}{2\omega_p}  e^{ip(x-y)-i\omega_p(t-t')} = 0.
\end{align}

Correspondingly, after coarse-graining, our effective constraint reads analogously 
\begin{align}
   - \langle \pi_n\pi_m\rangle + G^{-1}_{nm} \langle \phi_n\phi_m \rangle = 0\, ,
\end{align}
where all correlations are understood as equal-time, and, for this section, we changed the notation to 
\begin{align}
    M_{nm} \rightarrow G^{-1}_{nm} = \int_{\Lambda_a}^{\Lambda_a} \frac{\mathrm{d}p}{2\pi}(-p^2-m^2) e^{ipa(n-m)}.
\end{align}

\subsection{$\lambda\varphi^4$ theory: Mass renormalization at $\mathcal{O}(\lambda)$}
In the interacting case, the constraint receives an extra term
\begin{align}
    -\langle \pi_n\pi_m\rangle + G^{-1}_{nl} \langle \phi_l\phi_m \rangle +\frac{\tilde{\lambda}}{3!} \langle \phi_n\phi_m^3\rangle = 0\, ,
\end{align}
where here $\tilde{\lambda}$ is the ansatz for the coupling constant of the $\varphi^4$ interaction. At the same time, the correlation functions receive a correction due to the interaction. At leading order $\mathcal{O}(\lambda)$, the correction is given by a mass shift
\begin{align}
    m^2\rightarrow M^2 = m^2 + \frac{\lambda}{2}\langle \varphi^2_x\rangle  \, .
\end{align}
In one spatial dimension, the mass shift is logarithmically divergent. We thus set a fixed UV cutoff $\Lambda$ and we get
\begin{align}
    \langle \varphi^2_x\rangle  = \int^\Lambda\frac{\mathrm{d}^2p}{(2\pi)^2} \frac{1}{p^2+m^2} = \frac{\mathrm{log}\left(\frac{\Lambda^2+m^2}{m^2}\right)}{2\pi }\, .
\end{align}
by rotating the integral to Euclidean (imaginary) time. To obtain the constraint at leading order $\mathcal{O}(\lambda)$, we accordingly replace $m^2\rightarrow M^2$ in all correlation functions.

We consider furthermore the simplification
\begin{align}
    \frac{\tilde{\lambda}}{3!}\langle \phi_n\phi_m^3\rangle = \frac{\tilde{\lambda}}{2}\langle \phi_n\phi_m\rangle\langle\phi_m^2\rangle\, ,
\end{align}
which is valid up to higher orders in $\lambda$ (assuming that $\tilde{\lambda} = \mathcal{O}(\lambda)$ in agreement with the perturbative analysis).
This term can be absorbed into the inverse $G$, i.e.
\begin{align}
    G^{-1}_{nm} +\frac{\tilde{\lambda}}{2} \langle\phi_m^2\rangle= \int^{\Lambda_a} \frac{\mathrm{d}p}{2\pi}(-p^2-\tilde{m}^2 + \frac{\tilde{\lambda}}{2}\langle\phi_m^2\rangle ) e^{ipa(n-m)}.
\end{align}
where $\tilde{m}$ is here the ansatz for the bare mass. The constraint then yields
\begin{align}
\mathcal{C}_{nm} &= \int^{\Lambda_a}\frac{\mathrm{d}p}{2\pi}	\frac{(p^2+M^2) - (p^2+\tilde{m}^2 -\frac{\tilde{\lambda}}{2}\langle\phi_m^2\rangle )}{2\omega_p} e^{ipa(n-m)} \nonumber\\
&=\left(m^2-\tilde{m}^2+\frac{\lambda}{2}\langle \varphi^2_x\rangle  -\frac{\tilde{\lambda}}{2}\langle\phi_m^2\rangle\right) \nonumber\\
&\qquad \times\int^{\Lambda_a}\frac{\mathrm{d}p}{2\pi}	\frac{1}{2\omega_p} e^{ipa(n-m)}\, .
\end{align}
Specifically, after summation over the relative spatial index, averaging over the entire volume, and normalization by the mass, we get
\begin{align}
	\frac{1}{|m|V^2}\sum_{n,m}\mathcal{C}_{nm} = \frac{m^2-\tilde{m}^2+\frac{\lambda}{2}\langle \varphi^2_x\rangle  -\frac{\tilde{\lambda}}{2}\langle\phi_m^2\rangle}{4\pi m^2} \, .
\end{align}

The learned couplings are given by the solution to this constraint. In the present case, we obtain the result analytically, as
\begin{align}
    \tilde{m}^2 &= m^2 + \frac{\lambda}{2}\frac{\mathrm{log}\left(\frac{\Lambda^2+m^2}{m^2}\right)}{2\pi } -\frac{\lambda}{2}\langle\phi_m^2\rangle\nonumber \\
    &= M^2 - \frac{\lambda}{2}\langle\phi_m^2\rangle\, ,
\end{align}
where we have replaced $\tilde{\lambda}\rightarrow \lambda$ at leading order. Assuming $\Lambda\gg\Lambda_a$, we furthermore get
\begin{align}
    \langle \phi_m^2\rangle &= \int^\Lambda\int^{\Lambda_a}\frac{\mathrm{d}^2p}{(2\pi)^2} \frac{1}{p^2+m^2}\nonumber\\
   &= \frac{1}{2 \pi^2}\int^{\Lambda_a}\mathrm{d}p^1 \frac{1}{\omega_p}\mathrm{ArcTan}\left(\frac{\Lambda}{\omega_p}\right)\nonumber\\
   &\approx \frac{1}{2 \pi^2}\int^{\Lambda_a}\mathrm{d}p^1 \frac{1}{\omega_p}\left(\frac{\pi}{2}-\frac{\omega_p}{\Lambda}\right)\quad p,m\ll \Lambda \nonumber\\
   &= \frac{1}{2 \pi}\mathrm{arcsinh}\left(\frac{\Lambda_a}{m}\right) + \mathcal{O}(\Lambda_a/\Lambda)\nonumber\\
   &\approx \frac{1}{2 \pi}\log\left(\frac{\Lambda_a}{m}+\sqrt{\frac{\Lambda_a^2}{m^2}+1}\right)\nonumber\\
   &\rightarrow \frac{1}{4 \pi}\log\left(\Lambda_a^2\right)\quad \Lambda_a \gg m\, .
\end{align}
Hence, the solution for the learned mass is
\begin{align}
	\tilde{m}^2 &= m^2 + \frac{\lambda}{4\pi }\mathrm{log}\left(\frac{\Lambda^2+m^2}{\left[\Lambda_a+\sqrt{\Lambda_a^2+m^2}\right]^2}\right)\, .
\end{align}

\subsection{$\lambda\varphi^4$ theory: Coupling renormalization at $\mathcal{O}(\lambda^2)$}
To go beyond the mass renormalization and capture the running of the coupling $\lambda$, we consider constraint equations for four-point functions. The constraint equations read
\begin{align}
\label{eq:equal-time-four-point-constraint}
\mathcal{C} &=	-\langle \pi_{n_1}\phi_{n_2}\phi_{n_3}\pi_m\rangle+ \mathrm{perm.} + G^{-1}_{ml} \langle  \phi_{n_1}\phi_{n_2}\phi_{n_3}\phi_l \rangle \nonumber\\
&\quad + \frac{\tilde{\lambda}}{3!} \langle \phi_{n_1}\phi_{n_2}\phi_{n_3} \phi_m^3 \rangle \, .
\end{align}
To simulate the outcome of our HL protocol, we first relate the equal-time limit of the correlation functions with the diagrammatic expansion in terms of Feynman propagators in more detail. Specifically, the unequal-time constraint for four-point functions is given by
\begin{align}
	\mathcal{C} &= \int\mathrm{d}^2y\, G^{\phi,-1}_{xy} \langle\mathrm{T} \phi_{1} \phi_{2}\phi_{3}\phi_{y}\rangle + \frac{\tilde{\lambda}}{3!} \langle \mathrm{T} \phi_{1} \phi_{2}\phi_{3}\phi_{x}^3\rangle \nonumber\\
	&-\langle \mathrm{T} \phi_{1} \phi_{2}\rangle\delta(3-x) -\langle \mathrm{T} \phi_{1} \phi_{3}\rangle\delta(2-x) -\langle \mathrm{T} \phi_{2} \phi_{3}\rangle\delta(1-x) ,
\end{align}
where the indices now refer to space and time coordinates, and $G^{\phi}$ is the coarse-grained, time-ordered (Feynman) propagator for the fields $\phi$. Here, the second line is canceled by the disconnected components of the kinetic part of the constraint in accordance with the previous subsection. The remaining part to this order reads
\begin{align}
	\mathcal{C}^\mathrm{c} &= \int\mathrm{d}^2y\,    G^{\phi,-1}_{xy} \langle\mathrm{T} \phi_{1} \phi_{2}\phi_{3}\phi_{y}\rangle_\textbf{c} \nonumber\\
 &+ \frac{\tilde{\lambda}}{3!} \langle \mathrm{T} \phi_{1} \phi_{2}\phi_{x}^2\rangle \langle \mathrm{T}\phi_{3}\phi_{x}\rangle +\mathrm{perm} \, .
\end{align}
We notice here, that the role of the kinetic part, i.e. $ G^{\phi,-1} $, is to ``amputate" the external leg with index $y$ of all the diagrams in perturbation theory. Afterwards, all time-derivative terms are gone and we may safely take the equal-time limit. Then the diagrams in the first and the second term cancel piece by piece, up to perturbative corrections. These corrections originate in the fact, that for the first term the loop integral involves all momenta $|p|<\Lambda$, while in the second term they involve only momenta $|p|<\Lambda_a$. The missing part is absorbed into the shift of the learned couplings relative to the exact microscopic ones and constitutes the renormalization, similar to the previous subsection. 

We finally collecting all diagrams to that order and average over the position of the ``external fields''. Under the previously considered two-point constraint, the current four-point constraint then takes the form,
\begin{align}
\label{eq:four-constraint}
    \mathcal{C} &\propto 
    \frac{\lambda }{8m^3}\left(1-  \frac{3\lambda}{4\pi m^2}\sqrt{\frac{\Lambda^2}{\Lambda^2+m^2}}\right) \nonumber\\
    &- \frac{\tilde{\lambda} }{8m^3}\left(1-  \frac{3\lambda}{4\pi m^2}\sqrt{\frac{\Lambda_a^2}{\Lambda_a^2+m^2}}\right) \, ,
\end{align}
where the cutoff dependence originates from the loop expression
\begin{align}
    & \int^\Lambda\int^{\Lambda_a}\frac{\mathrm{d}^2p}{(2\pi)^2} \frac{1}{(p^2+m^2)^2}\nonumber\\
   &= \frac{1}{4 \pi^2}\int^{\Lambda_a}\mathrm{d}p \frac{1}{\omega_p^3}\mathrm{ArcTan}\left(\frac{\Lambda}{\omega_p}\right)+ \mathcal{O}(\Lambda_a/\Lambda)\nonumber\\
   &= \frac{1}{8 \pi}\int^{\Lambda_a}\mathrm{d}p \frac{1}{\omega_p^3} + \mathcal{O}(\Lambda_a/\Lambda)\nonumber\\
   &= \frac{1}{4 \pi m^2} \sqrt{\frac{\Lambda_a^2}{\Lambda_a^2+m^2}} + \mathcal{O}(\Lambda_a/\Lambda) \, ,
\end{align}
in the limit of $\Lambda_a\ll \Lambda$ for the second line, and
\begin{align}
    & \int^\Lambda\frac{\mathrm{d}^2p}{(2\pi)^2} \frac{1}{(p^2+m^2)^2}\nonumber\\
   &\approx \frac{1}{2\pi}\int^{\Lambda}\mathrm{d}p\, \frac{p}{\omega_p^4}\nonumber\\
   &= \frac{1}{4 \pi m^2} \sqrt{\frac{\Lambda^2}{\Lambda^2+m^2}} \, ,
\end{align}
for the first line of the constraint~\ref{eq:four-constraint}. Analytically, and in leading order, the constraint has the following solutions
\begin{align}
    \tilde{m}^2(a) &= m^2 + \frac{\lambda}{4\pi }\mathrm{log}\left(\frac{\Lambda^2+m^2}{\left[\Lambda_a+\sqrt{\Lambda_a^2+m^2}\right]^2}\right)\, ,\\
    \tilde{\lambda}(a)&=\lambda-  \lambda\frac{3\lambda }{4\pi m^2}\left(\sqrt{\frac{\Lambda^2}{\Lambda^2+m^2}}-\sqrt{\frac{\Lambda_a^2}{\Lambda_a^2+m^2}}\right)\, .
\end{align}
From this, we obtain the $\beta$ functions for the dimensionless couplings shown in the main text.

\section{Real-time dynamics of free theory}
\label{sec:quench}
In this section, we consider details of the Hamiltonian learning protocol with real-time evolving quantum fields following a mass quench of the lattice field theory. Suppose again that there is a cutoff $\Lambda$ for the lattice theory and a corresponding cutoff $\Lambda_a$ which describes the resolution scale. We consider the simple case of free field theories to illustrate the workings of our proposal. The free-field correlation functions can be obtained from the free-field expansion 
\begin{align}
	\varphi(x) &= \frac{1}{V}\sum_p \frac{1}{\sqrt{2\bar{\omega}_p}} e^{ipx}( a_p +  a_{-p}^\dagger)\, , \\
	\pi(x) &= \frac{1}{V}(-i)\sum_p \sqrt{\frac{\bar{\omega}_p}{2}}e^{ipx}( a_p -  a_{-p}^\dagger) \, ,
\end{align}
Specifically, this expansion has been chosen because then the Hamiltonian with mass $m$ is diagonal in the operators $a_p,a_{-p}^\dagger$. However, we may similarly define operators $b_p,b_{-p}^\dagger$ which diagonalize the Hamiltonian with a different mass $m'$. In a mass quench $m\rightarrow m'$ both of these descriptions come into play as we will detail below.
We first define the Fourier field operators as
\begin{align}
	\varphi_p &= \frac{1}{\sqrt{2\bar{\omega}_p}} \left(a_p+a^\dagger_{-p}\right)\, ,\\
	\pi_p &= (-i)\sqrt{\frac{\bar{\omega}_p}{2}}\left(a_p-a^\dagger_{-p}\right) \, .
\end{align}
As a next step, we rewrite them as
\begin{align}
	a_p &= \sqrt{\frac{\bar{\omega}_p}{2}} \varphi_p + i  \frac{1}{\sqrt{2\bar{\omega}_p}}\pi_p\nonumber\\
	&=  \frac{1}{2} \sqrt{\frac{\bar{\omega}_p}{\bar{\omega}'_p}} \left(b_p+b^\dagger_{-p}\right) + \frac{1}{2} \sqrt{\frac{\bar{\omega}'_p}{\bar{\omega}_p}} \left(b_p-b^\dagger_{-p}\right) \nonumber\\
	&= f_{+}(m,m') b_p + f_{-}(m,m') b_{-p}^\dagger\, ,
\end{align}
where 
\begin{align}
	f_{\pm}(m,m') = \frac{1}{2} \sqrt{\frac{\bar{\omega}_p}{\bar{\omega}'_p}} \pm \frac{1}{2} \sqrt{\frac{\bar{\omega}'_p}{\bar{\omega}_p}}\, ,
\end{align}
Altogether, we get
\begin{align}
	a_p 	&= f_{+}(m,m') b_p + f_{-}(m,m') b_{-p}^\dagger\, , \\
	a_{-p}^\dagger 	&= f_{-}(m,m') b_p + f_{+}(m,m') b_{-p}^\dagger\, ,
\end{align}
This is a Bogolyubov transformation with determinant $\mathrm{Det}= f_{+}^2 - f_{-}^2 = 1$.

Consider, the mass quench $m\rightarrow m'$, i.e. we start the evolution with Hamiltonian $m'$ in the ground state $\ket{0}$ of the Hamiltonian with $m$. Then, we are interested in the following quantities:
\begin{align}
		\bra{0}a^\dagger_pa_p\ket{0} &= 	\bra{0}(f_{p,-}b_{-p}+f_{p,+}b_p^\dagger)(f_{p,+} b_p + f_{p,-} b_{-p}^\dagger )\ket{0}\nonumber\\
		&= f^2_{p,-}\bra{0}b_{-p}  b_{-p}^\dagger\ket{0}\nonumber\\
		&= f^2_{p,-}\, ,
\end{align}
\begin{align}
	\bra{0}a_{-p}a_p\ket{0} &= 	\bra{0}(f_{-p,+} b_{-p} + f_{p,-} b_{p}^\dagger )(f_{p,+} b_p + f_{p,-} b_{-p}^\dagger )\ket{0}\nonumber\\
	&= f_{p,+}f_{p,-}\bra{0}b_{-p}  b_{-p}^\dagger\ket{0}\nonumber\\
	&=  f_{p,+}f_{p,-}\, ,
\end{align}
\begin{align}
	\bra{0}a^\dagger_{-p}a^\dagger_p\ket{0} &= 	\bra{0}(f_{p,-}b_{p}+f_{p,+}b_{-p}^\dagger)(f_{p,-}b_{-p}+f_{p,+}b_p^\dagger)\ket{0}\nonumber\\
	&= f_{p,-}f_{p,+}\bra{0}b_{p}  b_{p}^\dagger\ket{0}\nonumber\\
	&= f_{p,-}f_{p,+}\, ,
\end{align}
\begin{align}
	\bra{0}a_pa^\dagger_p\ket{0} &= 	\bra{0}(f_{p,+} b_p + f_{p,-} b_{-p}^\dagger )(f_{p,-}b_{-p}+f_{p,+}b_p^\dagger)\ket{0}\nonumber\\
	&= f^2_{p,+}\bra{0}b_{p}  b_{p}^\dagger\ket{0}\nonumber\\
	&= f^2_{p,+}\, .
\end{align}
With this, we will calculate the time evolution of the expectation value of a Hamiltonian ansatz. First, let us consider the time evolution of the field operators
\begin{align}
	\varphi_p(t) &= \frac{1}{\sqrt{2\bar{\omega}_p}} \left(a_p(t)+a^\dagger_{-p}(t)\right)\nonumber\\
	&= \frac{1}{\sqrt{2\bar{\omega}_p}} \left(a_pe^{i\bar{\omega}_pt}+a^\dagger_{-p}e^{-i\bar{\omega}_pt}\right)\nonumber\\
	&=  \frac{1}{\sqrt{2\bar{\omega}_p}} \Big[(f_{p,+}e^{i\bar{\omega}_pt}+f_{p,-}e^{-i\bar{\omega}_pt})  b_p \nonumber\\
	&\quad + (f_{p,-}e^{i\bar{\omega}_pt}+f_{p,+}e^{-i\bar{\omega}_pt}) b^\dagger_{-p} \Big] \, ,
 \end{align}
\begin{align}
	\pi_p(t) &=(-i)\sqrt{\frac{\bar{\omega}_p}{2}}\left(a_p(t)-a^\dagger_{-p}(t)\right) \nonumber\\
	&=(-i)\sqrt{\frac{\bar{\omega}_p}{2}}\left(a_pe^{i\bar{\omega}_pt}-a^\dagger_{-p}e^{-i\bar{\omega}_pt}\right) \nonumber\\
	&=(-i)\sqrt{\frac{\bar{\omega}_p}{2}}\Big[(f_{p,+}e^{i\bar{\omega}_pt}-f_{p,-}e^{-i\bar{\omega}_pt})  b_p \nonumber\\
	&\quad + (f_{p,-}e^{i\bar{\omega}_pt}-f_{p,+}e^{-i\bar{\omega}_pt}) b^\dagger_{-p} \Big] \, .
\end{align}
The Hamiltonian involves the following correlation functions:
\begin{align}
	\varphi_p(t)\varphi_{-p}(t) &= \frac{1}{2\bar{\omega}_p}\Big[ (f_{p,+}e^{i\bar{\omega}_pt}+f_{p,-}e^{-i\bar{\omega}_pt})^2  b_p b_{-p}+ \cdots \Big]\, .
\end{align}
In the final expectation value, the following term remains
\begin{align}
		\bra{0} \varphi_p(t)\varphi_{-p}(t)\ket{0} &= \frac{1}{2\bar{\omega}_p}\Big[ (f_{p,+}e^{i\bar{\omega}_pt}+f_{p,-}e^{-i\bar{\omega}_pt})   \nonumber\\
		&\quad \times (f_{p,-}e^{i\bar{\omega}_pt}+f_{p,+}e^{-i\bar{\omega}_pt}) \Big]\nonumber\\
		&=\frac{1}{2\bar{\omega}_p}\Big[ f_{+}f_{-}(e^{2i\bar{\omega}_pt}+e^{-2i\bar{\omega}_pt})\nonumber\\
		&\quad + (f_+^2+f_-^2)\Big]\, .
\end{align}
Similarly, the correlation function of the conjugate momenta reads
\begin{align}
	\bra{0} \pi_p(t)\pi_{-p}(t)\ket{0} &= \frac{\bar{\omega}_p}{2}\Big[ (f_{p,+}e^{i\bar{\omega}_pt}-f_{p,-}e^{-i\bar{\omega}_pt})   \nonumber\\
	&\quad \times (f_{p,-}e^{i\bar{\omega}_pt}-f_{p,+}e^{-i\bar{\omega}_pt})\Big]\nonumber\\
	&= \frac{\bar{\omega}_p}{2}\Big[ f_{+}f_{-}(e^{2i\bar{\omega}_pt}+e^{-2i\bar{\omega}_pt})\nonumber\\
	&\quad - (f_+^2+f_-^2)\Big]\, . 
\end{align}
After coarse-graining the observables, the (time-dependent) expectation value of the ansatz Hamiltonian in the continuum is then given by
\begin{align}
	\langle H(t)\rangle &= \int^{\Lambda_a}\frac{\mathrm{d}p}{(2\pi)}  \Big[\Big(-\frac{\bar{\omega}_p}{2}-\frac{\omega^2_p}{2\bar{\omega}_p}\Big)(f_+^2+f_-^2)\nonumber\\
	&+\Big(\frac{\bar{\omega}_p}{2}-\frac{\omega^2_p}{2\bar{\omega}_p}\Big)f_{+}f_{-}(e^{2i\bar{\omega}_pt}+e^{-2i\bar{\omega}_pt})\Big] \, .
\end{align}
From this, we get the constraint
\begin{align}
	\mathcal{C}_t& = 	\langle H(t)\rangle -	\langle H(0)\rangle \nonumber\\
	&=  \int^{\Lambda_a}\frac{\mathrm{d}p}{(2\pi)} \bar{\omega}_p\Big(1-\frac{\omega^2_p}{\bar{\omega}^2_p}\Big)f_{p,+}f_{p,-}[\cos(2\bar{\omega}_pt)-1]\, .
\end{align}
That means, at long times, the average constraint value is given by
\begin{align}
	\bar{\mathcal{C}} &= \frac{1}{T}\int_0^T\mathrm{d}t \, \mathcal{C}_t \rightarrow   \int^{\Lambda_a}\frac{\mathrm{d}p}{(2\pi)} \bar{\omega}_p\Big(\frac{\omega^2_p}{\bar{\omega}^2_p}-1\Big)f_{p,+}f_{p,-}\, .
\end{align}

\section{Classical-statistical field theory}
\label{sec:scale-dependent-classical}
\subsection{Classical limit of partition sum}
\label{sec:classical-limit}
\noindent In this section we derive the thermal equilibrium path integral in the classical (high-temperature) limit. We start with the quantum Hamiltonian $\hat{H} = \int_x [ \frac{1}{2}\hat{\pi}^2_x - \frac{1}{2}\hat{\varphi}_x \nabla^2 \hat{\varphi}_x + V(\hat{\varphi}_x) ]$ also given in the main text of the paper. Its quantum-thermal partition sum is given by
\begin{align}
    Z= \mathrm{Tr}\left( e^{-\beta \hat{H}}\right) = \int \mathcal{D}\phi\, e^{-S} \, .
\end{align}
Here $S$ is the microscopic, Euclidean action of the scalar fields
\begin{align}
S = \int \mathrm{d}x \int_0^\beta \mathrm{d}\tau \left[-\frac{1}{2} \varphi \,\partial^2 \varphi + V(\varphi)  \right]\, ,
\end{align}
where $\partial^2 = \partial_t^2+\nabla^2$ and the fields fulfill periodic boundary conditions along the imaginary time direction $\tau$, i.e.~$\varphi(x,\beta) = \varphi(x,0)$. In the high-temperature limit~($\beta\rightarrow0$), the path integral is dominated by the temporal zero mode $\varphi_0(x)= \varphi(x,\beta)$ with action
\begin{align}
    S \rightarrow \beta \int \mathrm{d}x \left[-\frac{1}{2} \varphi_0 \,\nabla^2 \varphi_0 + V(\varphi_0)  \right]\, ,
\end{align}
which yields the same result as starting from the classical Hamiltonian $H(\pi,\varphi)$ and integrating out the $\pi$ fields (upon renaming $\varphi_0\rightarrow \varphi$).

\subsection{Measurement model}
\label{sec:measurement-model}
\noindent In our experimental model, measurements are performed using a combination of optical elements/lenses and a CCD camera. We thus assume the measurement to include a Gaussian convolution of the fields through the optical lenses, followed by an integration over pixels of a camera, i.e. 
\begin{equation} W_{nx} = \int_{j-1/2}^{j+1/2}\frac{\mathrm{d}z}{a}  e^{-\frac{(z-x)^2}{2\sigma^2}} \, ,\end{equation} with resolution $\sigma$ and pixel width $a$. The remaining, unobservable, fields are given by higher "pixel Fourier modes", which we obtain via the coordinate transformation
\begin{equation}W_{n\alpha,x} = \int_{j-1/2}^{j+1/2}\frac{\mathrm{d}z}{a} e^{2\pi i \alpha \frac{z}{a}} e^{-\frac{(z-x)^2}{2\sigma^2}} \, ,\end{equation} with positive integers $\alpha \in \mathbb{N}_+$.

\subsection{Statistical analysis}
\label{sec:stat-analysis}
Assuming translation invariance, we compute constraints $\mathcal{C}^{(cl)}$ for local functionals $f[\phi] = f(\phi_m)$ as a function of $n-m$, and average over several values of the central coordinate $n+m$ to improve the statistical convergence. As the fluctuations of the constraint at different values of $n-m$ are correlated, we first transform the constraints to a frame in which they are statistically independent. To this end, we rotate $[\mathcal{C}^{(cl)}]_{k=n-m}$ to a new frame $[\mathcal{C}^{(cl)'}]_{k'} = \sum_k R_{k'k}[\mathcal{C}^{(cl)}]_{k}$, where the covariance matrix of the constraint data is diagonal in the different values of~$k'$, i.e. $\langle \mathcal{C}^{(cl)'}_{k'}\mathcal{C}^{(cl)'}_k\rangle \propto \delta_{kk'}$. In this frame, the constraints $\mathcal{C}^{(cl)'}_k$ are independent statistical variables, from which we compute $\chi^2/\nu $ as a model-independent quantifier. For a given ansatz, we then use a simple optimization routine to find the coupling values~$\boldsymbol{g}$ which minimize~$\chi^2/\nu$. 

\subsection{Effective Hamiltonian construction}
\noindent The experimentally accessible correlation functions are stored in the (high-temperature) path integral (with $\beta$ absorbed into the couplings). This may be written as
\begin{align}
    Z^{(cl)}[\bar{J}] = \int \mathcal{D}\varphi \, e^{-\frac{1}{2} \varphi \cdot G_0^{-1} \cdot \varphi - \int_x V(\varphi) + \bar{J} \cdot \bar{W} \varphi}\, .
\end{align}
Since $W$ is a basis change, it is invertible, and we may rewrite the path integral as
\begin{align}
    Z^{(cl)}[\bar{J}] &= \int \mathcal{D}\varphi \, e^{-\frac{1}{2}  \varphi_{\alpha i} W^{-1}_{x,\alpha i}  G^{-1}_{0,xy}  W^{-1}_{y,\beta j} \varphi_{\beta j}} \nonumber \\
    &\times e^{ - \int_x V(W^{-1}_{x,\alpha j} \varphi_{\alpha j}) + \bar{J} \phi} \, .
\end{align}
Here we introduced the notation of the main text, where $\phi_j = \varphi_{0j}$ is the field that is accessible in the measurement, and we imply summation/integration over repeated indices. To further ease notation, we define the new propagator
\begin{align}
    G^{-1}_{\alpha i,\beta j} = W^{-1}_{x,\alpha i}  G^{-1}_{0,xy}  W^{-1}_{y,\beta j} .
\end{align}
As a next step, we promote the path integral to include both the sources of observable and unobservable fields, bearing in mind that eventually only sources $\bar{J}_i$ are accessible, and adding corresponding sources $\tilde{J}_i = J_{\alpha>0,i}$ for convenience.
\begin{align}
    Z^{(cl)}[J] &= \int \mathcal{D}\varphi \, e^{-\frac{1}{2}  \varphi\, G^{-1}  \varphi - \int_x V(W^{-1} \varphi) + J  \pi} \nonumber\\
    &= e^{- \int_x V(W^{-1} [\delta/\delta J])}\int \mathcal{D}\varphi \, e^{-\frac{1}{2}  \varphi\, G^{-1}  \varphi  + J  \varphi}\, .
\end{align}

\subsubsection*{Free theory}
\noindent We first consider the free theory with $V=0$, i.e. the path integral is given by
\begin{align}
    Z_0^{(cl)}[J] = \int \mathcal{D}\varphi \, e^{-\frac{1}{2}  \varphi\, G^{-1}  \varphi  + J  \phi}\, .
\end{align}
We want to integrate out the inaccessible modes $\alpha > 0$. To achieve this, we first introduce an auxiliary field by inserting the following ``unity" into the path integral
\begin{align}
    1 = \int \mathcal{D}\phi\, \delta[\phi - \varphi ] = \int \mathcal{D}\phi\,\int \mathcal{D}\chi\, e^{i \chi (\phi - \varphi)} .
\end{align}
We get
\begin{align}
    Z_0^{(cl)}[J] = \int \mathcal{D}\phi \int \mathcal{D}\chi e^{i\chi \phi}\int \mathcal{D}\varphi\, e^{-\frac{1}{2}  \varphi\, G^{-1}  \varphi  + J  \phi}\, \Big|_{J = \mathcal{J}_{\chi} }.
\end{align}
where we defined
\begin{align}
    \mathcal{J}_{\chi} = \begin{pmatrix} \bar{J} +\chi \\ \tilde{J} \end{pmatrix}.
\end{align}
Subsequently, one finds
\begin{align}
    &Z_0^{(cl))}[J] = \int \mathcal{D}\phi \int \mathcal{D}\chi \,e^{i\chi \phi} e^{\frac{1}{2} \mathcal{J}_{\chi} G \mathcal{J}_{\chi} } \nonumber\\ 
    &=  \int \mathcal{D}\phi \int \mathcal{D}\chi \,e^{i\chi \phi} e^{\frac{1}{2} (\bar{J} + \chi) \bar{G} (\bar{J} + \chi) + \tilde{J} \tilde{G} \tilde{J} +[(\bar{J} + \chi) \bar{\tilde{G}} \tilde{J}  + C.C.] } .
\end{align}
where $C.C.$ abbreviates complex conjugation, and we introduced the following notation
\begin{align}
    G = \begin{pmatrix}
        \bar{G} && \bar{\tilde{G}} \\ \bar{\tilde{G}}^{T} && \tilde{G}
    \end{pmatrix} \, ,
\end{align}
where again ``bar" refers to the observable (``camera") modes, and ``tilde" to the modes that are integrated out. Performing the remaining $\chi$ integral yields
\begin{align}
    Z_0^{(cl)}[J] &= e^{\frac{1}{2} J G J}e^{-\frac{1}{2} J G (\bar{G})^{-1} G J} \nonumber\\
    &\times \int \mathcal{D}\phi \, e^{-\frac{1}{2} \phi (\bar{G})^{-1} \phi + \phi (\bar{G})^{-1} G J} , 
\end{align}
where here we emphasize that $(\bar{G})^{-1}$ is the inverse of the sub-matrix of $G$ referring to the camera modes (we will frequently drop the brackets in the following). When we omit the sources $\tilde{J}$ this expression simplifies to
\begin{align}
    Z_0^{(cl)}[\bar{J}] &= \int \mathcal{D}\phi \, e^{-\frac{1}{2} \phi (\bar{G})^{-1} \phi + \phi \bar{J}} , 
\end{align}
which, if differentiated with respect to sources $\bar{J}$, yields the correct convoluted and pixel-integrated correlations.

\subsubsection*{Interacting theory}
\noindent In the interacting case, we obtain
\begin{align}
    Z^{(cl)}[\bar{J}] &= e^{- \int_x V(W^{-1} \delta/\delta J)}Z^{(cl)}_0[J]\Big|_{\tilde{J}=0}\nonumber\\
    &=\int \mathcal{D}\phi \, e^{-\frac{1}{2} \phi (\bar{G})^{-1} \phi + \phi \bar{J}} e^{-V^{\mathrm{eff}}[\phi]}\, ,
\end{align}
i.e. an effective field theory with an effective potential~$V^{\mathrm{eff}}$, which consists of all connected diagrams assembled from the vertices of the classical potential~$V$ and the two types of propagators $G$ and $-G \bar{G}^{-1} G$. Note that $V^{\mathrm{eff}}$ does not contain further instances of $\bar{J}$ or $\tilde{J}$ since the latter are dropped after applying all the derivatives and the former cancel: if coupled to $\bar{J}$ the two propagators reduce to negatives of each other.

To make further progress, we assume that
\begin{align}
    \Delta = G-G \bar{G}^{-1} G\, ,
\end{align}
is a small quantity, which provides a controlled expansion parameter: The camera modes of $\Delta$ vanish, so $\Delta$ is only supported for higher pixel momenta $\alpha >0$. If the resolution $\sigma$ and the pixel size $a$ are small enough compared to the physical scales of the system, we expect these modes to yield only perturbative contributions in powers of $\sigma/\ell$ and $a/\ell$.

\noindent At leading order, we consider those diagrams without any $\Delta$ propagators, i.e.
\begin{align}
    \mathrm{LO:}\ V^{\mathrm{eff}}_\mathbf{LO} = \int_x V(W^{-1}_{x,\alpha j} G_{\alpha j, 0i} \bar{G}^{-1}_{i,n} \phi_n) .
\end{align}

\noindent The next order is given by all terms linear in $\Delta$. One can explicitly verify that this is given by 
\begin{align}
    \mathrm{NLO:}\ &V^{\mathrm{eff}}_\mathbf{NLO} =  V^{\mathrm{eff}}_\mathbf{LO} \nonumber\\
    & + \int_{x,y}  V'(W_{x,\cdot} G \bar{G}^{-1}\psi)   V'(W_{y,\cdot} G \bar{G}^{-1}\psi)\nonumber\\
    &\qquad \times  W_{x,\alpha i}\Delta_{\alpha i, \beta j} W_{y,\beta j} .
\end{align}

\subsection{Hamiltonian constraint for classical fields}
\label{sec:Schwinger-Dyson}
\noindent To apply Hamiltonian learning at a given pixel and resolution scale ($a$ and $\sigma$), we require a set of Hamiltonian constraints. That means, given a field-theoretic ansatz for the Hamiltonian, we compute constraints for the Hamiltonian couplings at the camera scale. Constraints may be derived from a path integral via the
Schwinger-Dyson equations,
\begin{align}
    Z^{(cl)}[\bar{J}] &=\int \mathcal{D}\phi \, e^{-\frac{1}{2} \phi (\bar{G})^{-1} \phi -V^{\mathrm{eff}}[\phi] + \phi \bar{J}}\nonumber\\
    &=\int \mathcal{D}\phi' \, e^{-\frac{1}{2} (\phi' + \epsilon)  (\bar{G})^{-1} (\phi' + \epsilon) -V^{\mathrm{eff}}[\phi' + \epsilon] + (\phi' + \epsilon) \bar{J}}\nonumber\\
    &\approx \int \mathcal{D}\phi' \, e^{-\frac{1}{2} \phi'  (\bar{G})^{-1} \phi' -V^{\mathrm{eff}}[\phi'] + \phi' \bar{J}}\nonumber\\ &\qquad \times \exp\left[-\epsilon \left(\bar{G}^{-1} \phi' + \frac{\delta V^{\mathrm{eff}}[\phi']}{\delta \phi'} - \bar{J}\right)\right]\nonumber\\
    &\approx \int \mathcal{D}\phi \, e^{-\frac{1}{2} \phi  (\bar{G})^{-1} \phi -V^{\mathrm{eff}}[\phi] + \phi \bar{J}}\nonumber\\ &\qquad \times \left[ 1-\epsilon \left(\bar{G}^{-1} \phi + \frac{\delta V^{\mathrm{eff}}[\phi]}{\delta \phi} - \bar{J}\right)\right]\, .
\end{align}
From this, we get
\begin{align}
      \left[ \bar{G}^{-1} \frac{\delta}{\delta \bar{J}} + \frac{\delta V^{\mathrm{eff}}[\phi]}{\delta \phi}\Big|_{\phi = \delta/\delta \bar{J}} - \bar{J} \right]Z^{(cl))}[\bar{J}] = 0 \, .
\end{align}
Higher-order constraints may subsequently derived by applying further functional derivatives to this equation. Note that all quantities are purely evaluated on camera pixel indices and are therefore experimentally accessible.

\section{Thermal Gaussian correlations}
\label{sec:thermal-Gaussian}
In this section, we compute the input data for the demonstration of HL in the quantum field case. For simplicity, we focus here on free thermal correlations, generated from a quadratic Hamiltonian. As before, we use a free field expansion and now consider the thermal state $\rho_\beta = e^{-\beta H}/\mathrm{Tr}[e^{-\beta H}]$, with occupation number
\begin{align}
    \langle a^\dagger_p a_q\rangle =  (2\pi)\delta(p-q)f_q .
\end{align}
Here, $f_q = (e^{\beta\omega_q} - 1)^{-1}$ is the Bose-Einstein distribution at temperature $1/\beta$. The thermal correlation functions are
\begin{align}
    \langle \varphi(x,t)\varphi(y,t') \rangle = \int \frac{\mathrm{d}p}{2\pi}\frac{1}{2\omega_p}&\bigg[  e^{ip(x-y)-i\omega_p(t-t')} (f_p+1)\nonumber\\
    &+ e^{-ip(x-y)+i\omega_p(t-t')} f_p\bigg]\ .
\end{align}
A "measurement'' introduces the additional resolution scale $a$ through the measurement kernel $W$, such that the measured field correlations are given by
\begin{align}
 &\langle \varphi(x,t)\varphi(y,t') \rangle\rightarrow \langle \phi_n(t)\phi_m(t') \rangle \nonumber\\
 &= \int^{\Lambda_a} \frac{\mathrm{d}p}{2\pi}\frac{1}{2\omega_p}\bigg[  e^{ipa(n-m)-i\omega_p(t-t')} (f_p+1)\nonumber\\
    &+ e^{-ip(x'-y')+i\omega_p(t-t')} f_p\bigg] .
\end{align}
Using again the two-point constraint for an ansatz Hamiltonian with mass $M$ the constraint will generally differ from zero. In the present case, the difference will be
\begin{align}
    \mathcal{C}(M^2) = (M^2-m^2) \langle \phi_n(t)\phi_m(t) \rangle \ .
\end{align}
The statistically independent components are the discrete momentum modes
\begin{align}
    &\sum_{n,m} e^{ik' n} e^{ikm} \langle \phi_n(t)\phi_m(t) \rangle \nonumber\\
    &= \int^{\Lambda_a} \frac{\mathrm{d}p}{2\pi}\frac{1}{2\omega_p}\bigg[   (f_p+1) \sum_{rs} \delta(k'+pa+2\pi r) \delta(k-pa+2\pi s)\nonumber\\
    &+ f_p \sum_{rs} \delta(k'-pa+2\pi r) \delta(k+pa+2\pi s)\bigg] \frac{\sin(pa/2)}{pa/2}. \nonumber\\
    &= \delta(k+k')\sum_{|r|,|s|< \Lambda_a a}\frac{1}{2\omega_{k+2\pi r}}\bigg[   (f_{k+2\pi r}+1)   \nonumber\\
    &\qquad \qquad \qquad \qquad+ f_{k+2\pi r} \bigg] \frac{\sin({k/2+\pi r})}{{k/2+\pi r}}\, .
\end{align}
This expression simplifies to
\begin{align}
    \sim \delta(k+k')\frac{f_{k}+\frac{1}{2}}{\omega_{k}}  \frac{\sin({k/2})}{{k/2}}\, .
\end{align}
In the Gaussian state, we can evaluate the sample complexity explicitly. We consider the case of $\nu+1$ measured modes, where $\nu$ is the remaining number of degrees of freedom for the one-parameter case of a massive free theory. We furthermore assume, that we measure correlation functions using a sample size $N$. For large sample sizes, we obtain mean values and variances as
\begin{align}
    \bar{\mu}_k &\sim \delta(k+k')(M^2-m^2)\frac{f_{k}+\frac{1}{2}}{\omega_{k}}  \frac{\sin({k/2})}{{k/2}} + x_k\bigg(\frac{ \bar{\sigma}_k}{\sqrt{N}} \bigg) ,\\
    \bar{\sigma}_k &\sim \delta(k+k')\frac{ (f_{k}+\frac{1}{2})}{\omega_{k}}  \frac{\sin({k/2})}{{k/2}} \sqrt{\omega^4_k + k^4 + m^4} ,
\end{align}
where $x_k$ is a Gaussian random variable with $\langle x_k \rangle = 0$ and $\langle x_k x_p\rangle = \delta_{kp}$. In total, we thus expect the following value for the reduced $\chi^2$
\begin{align}
    \chi^2/\nu &=\frac{1}{\nu} \sum_k \left[ \frac{\bar{\mu}_k}{\bar{\sigma}_k/\sqrt{N}} \right]^2\nonumber\\
    &\rightarrow \frac{ 1}{\nu} \sum_k \frac{((M^2 -m^2) + x_k  \sqrt{\omega^4_k + k^4 + m^4}/\sqrt{N})^2 }{(\omega^4_k+k^4+m^4) / N} \ .
\end{align}
The expression contains three parts:
\begin{align}
   (1):  \frac{ 1}{\nu} \sum_k \frac{(M^2 -m^2)^2 }{(\omega^4_k+k^4+m^4) / N}\, ,
\end{align}
\begin{align}
   (2):  \frac{ 1}{\nu} \sum_k 2\frac{(M^2 -m^2)x_k  \sqrt{\omega^4_k + k^4 + m^4}/\sqrt{N} }{(\omega^4_k+k^4+m^4) / N}\, ,
\end{align}

\begin{align}
   (3):  \frac{ 1}{\nu} \sum_k x^2_k \rightarrow 1 \, . 
\end{align}

For $M^2=m^2$, part (3) dominates and we expect $\chi^2/\nu \rightarrow 1$ for a large sample size. However, if $M^2\neq m^2$, one expects $\chi^2/\nu \sim \mathcal{O}(N)$ such that for sufficient sample sizes the ansatz will be discarded.

A similar analysis holds in the case of an interacting ansatz Hamiltonian, where we get
\begin{align}
    \mathcal{C}(\Delta\lambda) = \frac{\Delta\lambda}{3!} \langle \phi_n\phi^3_m \rangle = \frac{\Delta\lambda}{2!} \langle \phi_n\phi_m \rangle \langle \phi^2_m \rangle \ ,
\end{align}
using $\mathcal{C}(\Delta\lambda=0) =0$ for the "true" parameters as well as a decomposition of the four-point function into two-point correlations for the Gaussian data.
While the variances $\bar{\sigma}_k$ will again be dominated by the free fields, the mean values will change to 
\begin{align}
    \bar{\mu}_k \sim \delta(k+k')\frac{\Delta \lambda}{2!}\frac{f_{k}+\frac{1}{2}}{\omega_{k}}  \frac{\sin({k/2})}{{k/2}}\langle \phi^2\rangle + x_k\bigg(\frac{ \bar{\sigma}_k}{\sqrt{N}} \bigg)\, ,
\end{align}
and will hence depend on the temperature through the occupation number $f_k$. For the explicit plots shown in \Fig{fig:Statistics}, we numerically compute the above expressions for simulated Gaussian random numbers $x_k$ in the large $N$ approximation.

\clearpage
\bibliographystyle{apsrev4-1} 
\bibliography{Bib}

\begin{thebibliography}{52}%
\makeatletter
\providecommand \@ifxundefined [1]{%
 \@ifx{#1\undefined}
}%
\providecommand \@ifnum [1]{%
 \ifnum #1\expandafter \@firstoftwo
 \else \expandafter \@secondoftwo
 \fi
}%
\providecommand \@ifx [1]{%
 \ifx #1\expandafter \@firstoftwo
 \else \expandafter \@secondoftwo
 \fi
}%
\providecommand \natexlab [1]{#1}%
\providecommand \enquote  [1]{``#1''}%
\providecommand \bibnamefont  [1]{#1}%
\providecommand \bibfnamefont [1]{#1}%
\providecommand \citenamefont [1]{#1}%
\providecommand \href@noop [0]{\@secondoftwo}%
\providecommand \href [0]{\begingroup \@sanitize@url \@href}%
\providecommand \@href[1]{\@@startlink{#1}\@@href}%
\providecommand \@@href[1]{\endgroup#1\@@endlink}%
\providecommand \@sanitize@url [0]{\catcode `\\12\catcode `\$12\catcode
  `\&12\catcode `\#12\catcode `\^12\catcode `\_12\catcode `\%12\relax}%
\providecommand \@@startlink[1]{}%
\providecommand \@@endlink[0]{}%
\providecommand \url  [0]{\begingroup\@sanitize@url \@url }%
\providecommand \@url [1]{\endgroup\@href {#1}{\urlprefix }}%
\providecommand \urlprefix  [0]{URL }%
\providecommand \Eprint [0]{\href }%
\providecommand \doibase [0]{http://dx.doi.org/}%
\providecommand \selectlanguage [0]{\@gobble}%
\providecommand \bibinfo  [0]{\@secondoftwo}%
\providecommand \bibfield  [0]{\@secondoftwo}%
\providecommand \translation [1]{[#1]}%
\providecommand \BibitemOpen [0]{}%
\providecommand \bibitemStop [0]{}%
\providecommand \bibitemNoStop [0]{.\EOS\space}%
\providecommand \EOS [0]{\spacefactor3000\relax}%
\providecommand \BibitemShut  [1]{\csname bibitem#1\endcsname}%
\let\auto@bib@innerbib\@empty
\bibitem [{\citenamefont {National Academies~of Sciences}\ \emph
  {et~al.}(2020)\citenamefont {National Academies~of Sciences}, \citenamefont
  {Medicine} \emph {et~al.}}]{national2020manipulating}%
  \BibitemOpen
  \bibfield  {author} {\bibinfo {author} {\bibfnamefont {E.}~\bibnamefont
  {National Academies~of Sciences}}, \bibinfo {author} {\bibnamefont
  {Medicine}},  \emph {et~al.},\ }\href@noop {} {\emph {\bibinfo {title}
  {Manipulating Quantum Systems: An Assessment of Atomic, Molecular, and
  Optical Physics in the United States}}}\ (\bibinfo  {publisher} {National
  Academies Press},\ \bibinfo {year} {2020})\BibitemShut {NoStop}%
\bibitem [{\citenamefont {Gross}\ \emph {et~al.}(2023)\citenamefont {Gross},
  \citenamefont {Sevrin},\ and\ \citenamefont {Zoller}}]{gross2023physics}%
  \BibitemOpen
  \bibfield  {author} {\bibinfo {author} {\bibfnamefont {D.~J.}\ \bibnamefont
  {Gross}}, \bibinfo {author} {\bibfnamefont {A.}~\bibnamefont {Sevrin}}, \
  and\ \bibinfo {author} {\bibfnamefont {P.}~\bibnamefont {Zoller}},\
  }\href@noop {} {\emph {\bibinfo {title} {Physics Of Quantum Information,
  The-Proceedings Of The 28th Solvay Conference On Physics}}}\ (\bibinfo
  {publisher} {World Scientific},\ \bibinfo {year} {2023})\BibitemShut
  {NoStop}%
\bibitem [{\citenamefont {Daley}\ \emph {et~al.}(2022)\citenamefont {Daley},
  \citenamefont {Bloch}, \citenamefont {Kokail}, \citenamefont {Flannigan},
  \citenamefont {Pearson}, \citenamefont {Troyer},\ and\ \citenamefont
  {Zoller}}]{daley2022practical}%
  \BibitemOpen
  \bibfield  {author} {\bibinfo {author} {\bibfnamefont {A.~J.}\ \bibnamefont
  {Daley}}, \bibinfo {author} {\bibfnamefont {I.}~\bibnamefont {Bloch}},
  \bibinfo {author} {\bibfnamefont {C.}~\bibnamefont {Kokail}}, \bibinfo
  {author} {\bibfnamefont {S.}~\bibnamefont {Flannigan}}, \bibinfo {author}
  {\bibfnamefont {N.}~\bibnamefont {Pearson}}, \bibinfo {author} {\bibfnamefont
  {M.}~\bibnamefont {Troyer}}, \ and\ \bibinfo {author} {\bibfnamefont
  {P.}~\bibnamefont {Zoller}},\ }\href@noop {} {\bibfield  {journal} {\bibinfo
  {journal} {Nature}\ }\textbf {\bibinfo {volume} {607}},\ \bibinfo {pages}
  {667} (\bibinfo {year} {2022})}\BibitemShut {NoStop}%
\bibitem [{\citenamefont {Carrasco}\ \emph {et~al.}(2021)\citenamefont
  {Carrasco}, \citenamefont {Elben}, \citenamefont {Kokail}, \citenamefont
  {Kraus},\ and\ \citenamefont {Zoller}}]{carrasco2021theoretical}%
  \BibitemOpen
  \bibfield  {author} {\bibinfo {author} {\bibfnamefont {J.}~\bibnamefont
  {Carrasco}}, \bibinfo {author} {\bibfnamefont {A.}~\bibnamefont {Elben}},
  \bibinfo {author} {\bibfnamefont {C.}~\bibnamefont {Kokail}}, \bibinfo
  {author} {\bibfnamefont {B.}~\bibnamefont {Kraus}}, \ and\ \bibinfo {author}
  {\bibfnamefont {P.}~\bibnamefont {Zoller}},\ }\href@noop {} {\bibfield
  {journal} {\bibinfo  {journal} {PRX Quantum}\ }\textbf {\bibinfo {volume}
  {2}},\ \bibinfo {pages} {010102} (\bibinfo {year} {2021})}\BibitemShut
  {NoStop}%
\bibitem [{\citenamefont {Eisert}\ \emph {et~al.}(2020)\citenamefont {Eisert},
  \citenamefont {Hangleiter}, \citenamefont {Walk}, \citenamefont {Roth},
  \citenamefont {Markham}, \citenamefont {Parekh}, \citenamefont {Chabaud},\
  and\ \citenamefont {Kashefi}}]{eisert2020quantum}%
  \BibitemOpen
  \bibfield  {author} {\bibinfo {author} {\bibfnamefont {J.}~\bibnamefont
  {Eisert}}, \bibinfo {author} {\bibfnamefont {D.}~\bibnamefont {Hangleiter}},
  \bibinfo {author} {\bibfnamefont {N.}~\bibnamefont {Walk}}, \bibinfo {author}
  {\bibfnamefont {I.}~\bibnamefont {Roth}}, \bibinfo {author} {\bibfnamefont
  {D.}~\bibnamefont {Markham}}, \bibinfo {author} {\bibfnamefont
  {R.}~\bibnamefont {Parekh}}, \bibinfo {author} {\bibfnamefont
  {U.}~\bibnamefont {Chabaud}}, \ and\ \bibinfo {author} {\bibfnamefont
  {E.}~\bibnamefont {Kashefi}},\ }\href@noop {} {\bibfield  {journal} {\bibinfo
   {journal} {Nature Reviews Physics}\ }\textbf {\bibinfo {volume} {2}},\
  \bibinfo {pages} {382} (\bibinfo {year} {2020})}\BibitemShut {NoStop}%
\bibitem [{\citenamefont {Anshu}\ \emph {et~al.}(2021)\citenamefont {Anshu},
  \citenamefont {Arunachalam}, \citenamefont {Kuwahara},\ and\ \citenamefont
  {Soleimanifar}}]{anshu_sample-efficient_2021}%
  \BibitemOpen
  \bibfield  {author} {\bibinfo {author} {\bibfnamefont {A.}~\bibnamefont
  {Anshu}}, \bibinfo {author} {\bibfnamefont {S.}~\bibnamefont {Arunachalam}},
  \bibinfo {author} {\bibfnamefont {T.}~\bibnamefont {Kuwahara}}, \ and\
  \bibinfo {author} {\bibfnamefont {M.}~\bibnamefont {Soleimanifar}},\ }\href
  {\doibase 10.1038/s41567-021-01232-0} {\bibfield  {journal} {\bibinfo
  {journal} {Nature Physics}\ }\textbf {\bibinfo {volume} {17}},\ \bibinfo
  {pages} {931} (\bibinfo {year} {2021})}\BibitemShut {NoStop}%
\bibitem [{\citenamefont {Qi}\ and\ \citenamefont
  {Ranard}(2019)}]{qi2019determining}%
  \BibitemOpen
  \bibfield  {author} {\bibinfo {author} {\bibfnamefont {X.-L.}\ \bibnamefont
  {Qi}}\ and\ \bibinfo {author} {\bibfnamefont {D.}~\bibnamefont {Ranard}},\
  }\href@noop {} {\bibfield  {journal} {\bibinfo  {journal} {Quantum}\ }\textbf
  {\bibinfo {volume} {3}},\ \bibinfo {pages} {159} (\bibinfo {year}
  {2019})}\BibitemShut {NoStop}%
\bibitem [{\citenamefont {Bairey}\ \emph {et~al.}(2019)\citenamefont {Bairey},
  \citenamefont {Arad},\ and\ \citenamefont {Lindner}}]{bairey_learning_2019}%
  \BibitemOpen
  \bibfield  {author} {\bibinfo {author} {\bibfnamefont {E.}~\bibnamefont
  {Bairey}}, \bibinfo {author} {\bibfnamefont {I.}~\bibnamefont {Arad}}, \ and\
  \bibinfo {author} {\bibfnamefont {N.~H.}\ \bibnamefont {Lindner}},\ }\href
  {\doibase 10.1103/PhysRevLett.122.020504} {\bibfield  {journal} {\bibinfo
  {journal} {Physical Review Letters}\ }\textbf {\bibinfo {volume} {122}},\
  \bibinfo {pages} {020504} (\bibinfo {year} {2019})}\BibitemShut {NoStop}%
\bibitem [{\citenamefont {Li}\ \emph {et~al.}(2020)\citenamefont {Li},
  \citenamefont {Zou},\ and\ \citenamefont {Hsieh}}]{li_hamiltonian_2020}%
  \BibitemOpen
  \bibfield  {author} {\bibinfo {author} {\bibfnamefont {Z.}~\bibnamefont
  {Li}}, \bibinfo {author} {\bibfnamefont {L.}~\bibnamefont {Zou}}, \ and\
  \bibinfo {author} {\bibfnamefont {T.~H.}\ \bibnamefont {Hsieh}},\ }\href
  {\doibase 10.1103/PhysRevLett.124.160502} {\bibfield  {journal} {\bibinfo
  {journal} {Physical Review Letters}\ }\textbf {\bibinfo {volume} {124}},\
  \bibinfo {pages} {160502} (\bibinfo {year} {2020})}\BibitemShut {NoStop}%
\bibitem [{\citenamefont {Wiebe}\ \emph {et~al.}(2014)\citenamefont {Wiebe},
  \citenamefont {Granade}, \citenamefont {Ferrie},\ and\ \citenamefont
  {Cory}}]{wiebe2014hamiltonian}%
  \BibitemOpen
  \bibfield  {author} {\bibinfo {author} {\bibfnamefont {N.}~\bibnamefont
  {Wiebe}}, \bibinfo {author} {\bibfnamefont {C.}~\bibnamefont {Granade}},
  \bibinfo {author} {\bibfnamefont {C.}~\bibnamefont {Ferrie}}, \ and\ \bibinfo
  {author} {\bibfnamefont {D.~G.}\ \bibnamefont {Cory}},\ }\href {\doibase
  https://doi.org/10.1103/PhysRevLett.112.190501} {\bibfield  {journal}
  {\bibinfo  {journal} {{Physical Review Letters}}\ }\textbf {\bibinfo {volume}
  {112}},\ \bibinfo {pages} {190501} (\bibinfo {year} {2014})}\BibitemShut
  {NoStop}%
\bibitem [{\citenamefont {Valenti}\ \emph {et~al.}(2022)\citenamefont
  {Valenti}, \citenamefont {Jin}, \citenamefont {L{\'e}onard}, \citenamefont
  {Huber},\ and\ \citenamefont {Greplova}}]{valenti2022scalable}%
  \BibitemOpen
  \bibfield  {author} {\bibinfo {author} {\bibfnamefont {A.}~\bibnamefont
  {Valenti}}, \bibinfo {author} {\bibfnamefont {G.}~\bibnamefont {Jin}},
  \bibinfo {author} {\bibfnamefont {J.}~\bibnamefont {L{\'e}onard}}, \bibinfo
  {author} {\bibfnamefont {S.~D.}\ \bibnamefont {Huber}}, \ and\ \bibinfo
  {author} {\bibfnamefont {E.}~\bibnamefont {Greplova}},\ }\href {\doibase
  https://doi.org/10.1103/PhysRevA.105.023302} {\bibfield  {journal} {\bibinfo
  {journal} {Physical Review A}\ }\textbf {\bibinfo {volume} {105}},\ \bibinfo
  {pages} {023302} (\bibinfo {year} {2022})}\BibitemShut {NoStop}%
\bibitem [{\citenamefont {Pastori}\ \emph {et~al.}(2022)\citenamefont
  {Pastori}, \citenamefont {Olsacher}, \citenamefont {Kokail},\ and\
  \citenamefont {Zoller}}]{pastori2022characterization}%
  \BibitemOpen
  \bibfield  {author} {\bibinfo {author} {\bibfnamefont {L.}~\bibnamefont
  {Pastori}}, \bibinfo {author} {\bibfnamefont {T.}~\bibnamefont {Olsacher}},
  \bibinfo {author} {\bibfnamefont {C.}~\bibnamefont {Kokail}}, \ and\ \bibinfo
  {author} {\bibfnamefont {P.}~\bibnamefont {Zoller}},\ }\href {\doibase
  https://doi.org/10.1103/PRXQuantum.3.030324} {\bibfield  {journal} {\bibinfo
  {journal} {PRX Quantum}\ }\textbf {\bibinfo {volume} {3}},\ \bibinfo {pages}
  {030324} (\bibinfo {year} {2022})}\BibitemShut {NoStop}%
\bibitem [{\citenamefont {Shabani}\ \emph {et~al.}(2011)\citenamefont
  {Shabani}, \citenamefont {Mohseni}, \citenamefont {Lloyd}, \citenamefont
  {Kosut},\ and\ \citenamefont {Rabitz}}]{shabani2011estimation}%
  \BibitemOpen
  \bibfield  {author} {\bibinfo {author} {\bibfnamefont {A.}~\bibnamefont
  {Shabani}}, \bibinfo {author} {\bibfnamefont {M.}~\bibnamefont {Mohseni}},
  \bibinfo {author} {\bibfnamefont {S.}~\bibnamefont {Lloyd}}, \bibinfo
  {author} {\bibfnamefont {R.~L.}\ \bibnamefont {Kosut}}, \ and\ \bibinfo
  {author} {\bibfnamefont {H.}~\bibnamefont {Rabitz}},\ }\href {\doibase
  https://doi.org/10.1103/PhysRevA.84.012107} {\bibfield  {journal} {\bibinfo
  {journal} {Physical Review A}\ }\textbf {\bibinfo {volume} {84}},\ \bibinfo
  {pages} {012107} (\bibinfo {year} {2011})}\BibitemShut {NoStop}%
\bibitem [{\citenamefont {Browaeys}\ and\ \citenamefont
  {Lahaye}(2020)}]{browaeys2020many}%
  \BibitemOpen
  \bibfield  {author} {\bibinfo {author} {\bibfnamefont {A.}~\bibnamefont
  {Browaeys}}\ and\ \bibinfo {author} {\bibfnamefont {T.}~\bibnamefont
  {Lahaye}},\ }\href@noop {} {\bibfield  {journal} {\bibinfo  {journal} {Nature
  Physics}\ }\textbf {\bibinfo {volume} {16}},\ \bibinfo {pages} {132}
  (\bibinfo {year} {2020})}\BibitemShut {NoStop}%
\bibitem [{\citenamefont {Altman}\ \emph {et~al.}(2021)\citenamefont {Altman},
  \citenamefont {Brown}, \citenamefont {Carleo}, \citenamefont {Carr},
  \citenamefont {Demler}, \citenamefont {Chin}, \citenamefont {DeMarco},
  \citenamefont {Economou}, \citenamefont {Eriksson}, \citenamefont {Fu} \emph
  {et~al.}}]{altman2021quantum}%
  \BibitemOpen
  \bibfield  {author} {\bibinfo {author} {\bibfnamefont {E.}~\bibnamefont
  {Altman}}, \bibinfo {author} {\bibfnamefont {K.~R.}\ \bibnamefont {Brown}},
  \bibinfo {author} {\bibfnamefont {G.}~\bibnamefont {Carleo}}, \bibinfo
  {author} {\bibfnamefont {L.~D.}\ \bibnamefont {Carr}}, \bibinfo {author}
  {\bibfnamefont {E.}~\bibnamefont {Demler}}, \bibinfo {author} {\bibfnamefont
  {C.}~\bibnamefont {Chin}}, \bibinfo {author} {\bibfnamefont {B.}~\bibnamefont
  {DeMarco}}, \bibinfo {author} {\bibfnamefont {S.~E.}\ \bibnamefont
  {Economou}}, \bibinfo {author} {\bibfnamefont {M.~A.}\ \bibnamefont
  {Eriksson}}, \bibinfo {author} {\bibfnamefont {K.-M.~C.}\ \bibnamefont {Fu}},
   \emph {et~al.},\ }\href {\doibase
  https://doi.org/10.1103/PRXQuantum.2.017003} {\bibfield  {journal} {\bibinfo
  {journal} {PRX Quantum}\ }\textbf {\bibinfo {volume} {2}},\ \bibinfo {pages}
  {017003} (\bibinfo {year} {2021})}\BibitemShut {NoStop}%
\bibitem [{\citenamefont {Kuhr}(2016)}]{kuhr2016quantum}%
  \BibitemOpen
  \bibfield  {author} {\bibinfo {author} {\bibfnamefont {S.}~\bibnamefont
  {Kuhr}},\ }\href@noop {} {\bibfield  {journal} {\bibinfo  {journal} {National
  Science Review}\ }\textbf {\bibinfo {volume} {3}},\ \bibinfo {pages} {170}
  (\bibinfo {year} {2016})}\BibitemShut {NoStop}%
\bibitem [{\citenamefont {Weinberg}(1995)}]{weinberg1995quantum}%
  \BibitemOpen
  \bibfield  {author} {\bibinfo {author} {\bibfnamefont {S.}~\bibnamefont
  {Weinberg}},\ }\href@noop {} {\emph {\bibinfo {title} {The quantum theory of
  fields}}}\ (\bibinfo  {publisher} {Cambridge university press},\ \bibinfo
  {year} {1995})\BibitemShut {NoStop}%
\bibitem [{\citenamefont {Altland}\ and\ \citenamefont
  {Simons}(2010)}]{altland2010condensed}%
  \BibitemOpen
  \bibfield  {author} {\bibinfo {author} {\bibfnamefont {A.}~\bibnamefont
  {Altland}}\ and\ \bibinfo {author} {\bibfnamefont {B.~D.}\ \bibnamefont
  {Simons}},\ }\href@noop {} {\emph {\bibinfo {title} {Condensed matter field
  theory}}}\ (\bibinfo  {publisher} {Cambridge university press},\ \bibinfo
  {year} {2010})\BibitemShut {NoStop}%
\bibitem [{\citenamefont {Sachdev}(2023)}]{sachdev2023quantum}%
  \BibitemOpen
  \bibfield  {author} {\bibinfo {author} {\bibfnamefont {S.}~\bibnamefont
  {Sachdev}},\ }\href@noop {} {\emph {\bibinfo {title} {Quantum Phases of
  Matter}}}\ (\bibinfo  {publisher} {Cambridge University Press},\ \bibinfo
  {year} {2023})\BibitemShut {NoStop}%
\bibitem [{\citenamefont {Fradkin}(2013)}]{fradkin2013field}%
  \BibitemOpen
  \bibfield  {author} {\bibinfo {author} {\bibfnamefont {E.}~\bibnamefont
  {Fradkin}},\ }\href@noop {} {\emph {\bibinfo {title} {Field theories of
  condensed matter physics}}}\ (\bibinfo  {publisher} {Cambridge University
  Press},\ \bibinfo {year} {2013})\BibitemShut {NoStop}%
\bibitem [{\citenamefont {Gardiner}\ and\ \citenamefont
  {Zoller}(2017)}]{gardiner2017quantum}%
  \BibitemOpen
  \bibfield  {author} {\bibinfo {author} {\bibfnamefont {C.~W.}\ \bibnamefont
  {Gardiner}}\ and\ \bibinfo {author} {\bibfnamefont {P.}~\bibnamefont
  {Zoller}},\ }\href@noop {} {\emph {\bibinfo {title} {Quantum World Of
  Ultra-cold Atoms And Light, The-Book Iii: Ultra-cold Atoms}}},\ Vol.~\bibinfo
  {volume} {5}\ (\bibinfo  {publisher} {World Scientific},\ \bibinfo {year}
  {2017})\BibitemShut {NoStop}%
\bibitem [{\citenamefont {Fisher}(1998)}]{fisher1998renormalization}%
  \BibitemOpen
  \bibfield  {author} {\bibinfo {author} {\bibfnamefont {M.~E.}\ \bibnamefont
  {Fisher}},\ }\href@noop {} {\bibfield  {journal} {\bibinfo  {journal}
  {Reviews of Modern Physics}\ }\textbf {\bibinfo {volume} {70}},\ \bibinfo
  {pages} {653} (\bibinfo {year} {1998})}\BibitemShut {NoStop}%
\bibitem [{\citenamefont {Schweigler}\ \emph {et~al.}(2017)\citenamefont
  {Schweigler}, \citenamefont {Kasper}, \citenamefont {Erne}, \citenamefont
  {Mazets}, \citenamefont {Rauer}, \citenamefont {Cataldini}, \citenamefont
  {Langen}, \citenamefont {Gasenzer}, \citenamefont {Berges},\ and\
  \citenamefont {Schmiedmayer}}]{schweigler2017experimental}%
  \BibitemOpen
  \bibfield  {author} {\bibinfo {author} {\bibfnamefont {T.}~\bibnamefont
  {Schweigler}}, \bibinfo {author} {\bibfnamefont {V.}~\bibnamefont {Kasper}},
  \bibinfo {author} {\bibfnamefont {S.}~\bibnamefont {Erne}}, \bibinfo {author}
  {\bibfnamefont {I.}~\bibnamefont {Mazets}}, \bibinfo {author} {\bibfnamefont
  {B.}~\bibnamefont {Rauer}}, \bibinfo {author} {\bibfnamefont
  {F.}~\bibnamefont {Cataldini}}, \bibinfo {author} {\bibfnamefont
  {T.}~\bibnamefont {Langen}}, \bibinfo {author} {\bibfnamefont
  {T.}~\bibnamefont {Gasenzer}}, \bibinfo {author} {\bibfnamefont
  {J.}~\bibnamefont {Berges}}, \ and\ \bibinfo {author} {\bibfnamefont
  {J.}~\bibnamefont {Schmiedmayer}},\ }\href {\doibase
  https://doi.org/10.1038/nature22310} {\bibfield  {journal} {\bibinfo
  {journal} {Nature}\ }\textbf {\bibinfo {volume} {545}},\ \bibinfo {pages}
  {323} (\bibinfo {year} {2017})}\BibitemShut {NoStop}%
\bibitem [{\citenamefont {Pigneur}\ \emph {et~al.}(2018)\citenamefont
  {Pigneur}, \citenamefont {Berrada}, \citenamefont {Bonneau}, \citenamefont
  {Schumm}, \citenamefont {Demler},\ and\ \citenamefont
  {Schmiedmayer}}]{pigneur2018relaxation}%
  \BibitemOpen
  \bibfield  {author} {\bibinfo {author} {\bibfnamefont {M.}~\bibnamefont
  {Pigneur}}, \bibinfo {author} {\bibfnamefont {T.}~\bibnamefont {Berrada}},
  \bibinfo {author} {\bibfnamefont {M.}~\bibnamefont {Bonneau}}, \bibinfo
  {author} {\bibfnamefont {T.}~\bibnamefont {Schumm}}, \bibinfo {author}
  {\bibfnamefont {E.}~\bibnamefont {Demler}}, \ and\ \bibinfo {author}
  {\bibfnamefont {J.}~\bibnamefont {Schmiedmayer}},\ }\href {\doibase
  https://doi.org/10.1103/PhysRevLett.120.173601} {\bibfield  {journal}
  {\bibinfo  {journal} {Physical Review Letters}\ }\textbf {\bibinfo {volume}
  {120}},\ \bibinfo {pages} {173601} (\bibinfo {year} {2018})}\BibitemShut
  {NoStop}%
\bibitem [{\citenamefont {Tajik}\ \emph {et~al.}(2023)\citenamefont {Tajik},
  \citenamefont {Kukuljan}, \citenamefont {Sotiriadis}, \citenamefont {Rauer},
  \citenamefont {Schweigler}, \citenamefont {Cataldini}, \citenamefont
  {Sabino}, \citenamefont {M{\o}ller}, \citenamefont {Sch{\"u}ttelkopf},
  \citenamefont {Ji} \emph {et~al.}}]{tajik2023verification}%
  \BibitemOpen
  \bibfield  {author} {\bibinfo {author} {\bibfnamefont {M.}~\bibnamefont
  {Tajik}}, \bibinfo {author} {\bibfnamefont {I.}~\bibnamefont {Kukuljan}},
  \bibinfo {author} {\bibfnamefont {S.}~\bibnamefont {Sotiriadis}}, \bibinfo
  {author} {\bibfnamefont {B.}~\bibnamefont {Rauer}}, \bibinfo {author}
  {\bibfnamefont {T.}~\bibnamefont {Schweigler}}, \bibinfo {author}
  {\bibfnamefont {F.}~\bibnamefont {Cataldini}}, \bibinfo {author}
  {\bibfnamefont {J.}~\bibnamefont {Sabino}}, \bibinfo {author} {\bibfnamefont
  {F.}~\bibnamefont {M{\o}ller}}, \bibinfo {author} {\bibfnamefont
  {P.}~\bibnamefont {Sch{\"u}ttelkopf}}, \bibinfo {author} {\bibfnamefont
  {S.-C.}\ \bibnamefont {Ji}},  \emph {et~al.},\ }\href {\doibase
  https://doi.org/10.1038/s41567-023-02027-1} {\bibfield  {journal} {\bibinfo
  {journal} {Nature Physics}\ ,\ \bibinfo {pages} {1}} (\bibinfo {year}
  {2023})}\BibitemShut {NoStop}%
\bibitem [{\citenamefont {Gritsev}\ \emph {et~al.}(2007)\citenamefont
  {Gritsev}, \citenamefont {Polkovnikov},\ and\ \citenamefont
  {Demler}}]{gritsev2007linear}%
  \BibitemOpen
  \bibfield  {author} {\bibinfo {author} {\bibfnamefont {V.}~\bibnamefont
  {Gritsev}}, \bibinfo {author} {\bibfnamefont {A.}~\bibnamefont
  {Polkovnikov}}, \ and\ \bibinfo {author} {\bibfnamefont {E.}~\bibnamefont
  {Demler}},\ }\href {\doibase https://doi.org/10.1103/PhysRevB.75.174511}
  {\bibfield  {journal} {\bibinfo  {journal} {Physical Review B}\ }\textbf
  {\bibinfo {volume} {75}},\ \bibinfo {pages} {174511} (\bibinfo {year}
  {2007})}\BibitemShut {NoStop}%
\bibitem [{\citenamefont {Dalla~Torre}\ \emph {et~al.}(2013)\citenamefont
  {Dalla~Torre}, \citenamefont {Demler},\ and\ \citenamefont
  {Polkovnikov}}]{dalla2013universal}%
  \BibitemOpen
  \bibfield  {author} {\bibinfo {author} {\bibfnamefont {E.~G.}\ \bibnamefont
  {Dalla~Torre}}, \bibinfo {author} {\bibfnamefont {E.}~\bibnamefont {Demler}},
  \ and\ \bibinfo {author} {\bibfnamefont {A.}~\bibnamefont {Polkovnikov}},\
  }\href@noop {} {\bibfield  {journal} {\bibinfo  {journal} {Physical review
  letters}\ }\textbf {\bibinfo {volume} {110}},\ \bibinfo {pages} {090404}
  (\bibinfo {year} {2013})}\BibitemShut {NoStop}%
\bibitem [{\citenamefont {Imambekov}\ \emph {et~al.}(2008)\citenamefont
  {Imambekov}, \citenamefont {Gritsev},\ and\ \citenamefont
  {Demler}}]{imambekov2008mapping}%
  \BibitemOpen
  \bibfield  {author} {\bibinfo {author} {\bibfnamefont {A.}~\bibnamefont
  {Imambekov}}, \bibinfo {author} {\bibfnamefont {V.}~\bibnamefont {Gritsev}},
  \ and\ \bibinfo {author} {\bibfnamefont {E.}~\bibnamefont {Demler}},\
  }\href@noop {} {\bibfield  {journal} {\bibinfo  {journal} {Physical Review
  A}\ }\textbf {\bibinfo {volume} {77}},\ \bibinfo {pages} {063606} (\bibinfo
  {year} {2008})}\BibitemShut {NoStop}%
\bibitem [{\citenamefont {Giamarchi}(2003)}]{giamarchi2003quantum}%
  \BibitemOpen
  \bibfield  {author} {\bibinfo {author} {\bibfnamefont {T.}~\bibnamefont
  {Giamarchi}},\ }\href@noop {} {\emph {\bibinfo {title} {Quantum physics in
  one dimension}}},\ Vol.\ \bibinfo {volume} {121}\ (\bibinfo  {publisher}
  {Clarendon press},\ \bibinfo {year} {2003})\BibitemShut {NoStop}%
\bibitem [{\citenamefont {Cuevas-Maraver}\ \emph {et~al.}(2014)\citenamefont
  {Cuevas-Maraver}, \citenamefont {Kevrekidis},\ and\ \citenamefont
  {Williams}}]{cuevas2014sine}%
  \BibitemOpen
  \bibfield  {author} {\bibinfo {author} {\bibfnamefont {J.}~\bibnamefont
  {Cuevas-Maraver}}, \bibinfo {author} {\bibfnamefont {P.~G.}\ \bibnamefont
  {Kevrekidis}}, \ and\ \bibinfo {author} {\bibfnamefont {F.}~\bibnamefont
  {Williams}},\ }\href@noop {} {\bibfield  {journal} {\bibinfo  {journal}
  {Nonlinear systems and complexity}\ }\textbf {\bibinfo {volume} {10}}
  (\bibinfo {year} {2014})}\BibitemShut {NoStop}%
\bibitem [{\citenamefont {Kogut}(1979)}]{kogut1979introduction}%
  \BibitemOpen
  \bibfield  {author} {\bibinfo {author} {\bibfnamefont {J.~B.}\ \bibnamefont
  {Kogut}},\ }\href {\doibase https://doi.org/10.1103/RevModPhys.51.659}
  {\bibfield  {journal} {\bibinfo  {journal} {Reviews of Modern Physics}\
  }\textbf {\bibinfo {volume} {51}},\ \bibinfo {pages} {659} (\bibinfo {year}
  {1979})}\BibitemShut {NoStop}%
\bibitem [{\citenamefont {Coleman}(1976)}]{coleman1976more}%
  \BibitemOpen
  \bibfield  {author} {\bibinfo {author} {\bibfnamefont {S.}~\bibnamefont
  {Coleman}},\ }\href {\doibase https://doi.org/10.1016/0003-4916(76)90280-3}
  {\bibfield  {journal} {\bibinfo  {journal} {Annals of Physics}\ }\textbf
  {\bibinfo {volume} {101}},\ \bibinfo {pages} {239} (\bibinfo {year}
  {1976})}\BibitemShut {NoStop}%
\bibitem [{\citenamefont {Bücker}\ \emph {et~al.}(2009)\citenamefont
  {Bücker}, \citenamefont {Perrin}, \citenamefont {Manz}, \citenamefont
  {Betz}, \citenamefont {Koller}, \citenamefont {Plisson}, \citenamefont
  {Rottmann}, \citenamefont {Schumm},\ and\ \citenamefont
  {Schmiedmayer}}]{Buecker_2009}%
  \BibitemOpen
  \bibfield  {author} {\bibinfo {author} {\bibfnamefont {R.}~\bibnamefont
  {Bücker}}, \bibinfo {author} {\bibfnamefont {A.}~\bibnamefont {Perrin}},
  \bibinfo {author} {\bibfnamefont {S.}~\bibnamefont {Manz}}, \bibinfo {author}
  {\bibfnamefont {T.}~\bibnamefont {Betz}}, \bibinfo {author} {\bibfnamefont
  {C.}~\bibnamefont {Koller}}, \bibinfo {author} {\bibfnamefont
  {T.}~\bibnamefont {Plisson}}, \bibinfo {author} {\bibfnamefont
  {J.}~\bibnamefont {Rottmann}}, \bibinfo {author} {\bibfnamefont
  {T.}~\bibnamefont {Schumm}}, \ and\ \bibinfo {author} {\bibfnamefont
  {J.}~\bibnamefont {Schmiedmayer}},\ }\href {\doibase
  10.1088/1367-2630/11/10/103039} {\bibfield  {journal} {\bibinfo  {journal}
  {New Journal of Physics}\ }\textbf {\bibinfo {volume} {11}},\ \bibinfo
  {pages} {103039} (\bibinfo {year} {2009})}\BibitemShut {NoStop}%
\bibitem [{\citenamefont {Bakr}\ \emph {et~al.}(2009)\citenamefont {Bakr},
  \citenamefont {Gillen}, \citenamefont {Peng}, \citenamefont {F{\"o}lling},\
  and\ \citenamefont {Greiner}}]{bakr2009quantum}%
  \BibitemOpen
  \bibfield  {author} {\bibinfo {author} {\bibfnamefont {W.~S.}\ \bibnamefont
  {Bakr}}, \bibinfo {author} {\bibfnamefont {J.~I.}\ \bibnamefont {Gillen}},
  \bibinfo {author} {\bibfnamefont {A.}~\bibnamefont {Peng}}, \bibinfo {author}
  {\bibfnamefont {S.}~\bibnamefont {F{\"o}lling}}, \ and\ \bibinfo {author}
  {\bibfnamefont {M.}~\bibnamefont {Greiner}},\ }\href@noop {} {\bibfield
  {journal} {\bibinfo  {journal} {Nature}\ }\textbf {\bibinfo {volume} {462}},\
  \bibinfo {pages} {74} (\bibinfo {year} {2009})}\BibitemShut {NoStop}%
\bibitem [{\citenamefont {Langen}\ \emph {et~al.}(2015)\citenamefont {Langen},
  \citenamefont {Erne}, \citenamefont {Geiger}, \citenamefont {Rauer},
  \citenamefont {Schweigler}, \citenamefont {Kuhnert}, \citenamefont
  {Rohringer}, \citenamefont {Mazets}, \citenamefont {Gasenzer},\ and\
  \citenamefont {Schmiedmayer}}]{langen2015experimental}%
  \BibitemOpen
  \bibfield  {author} {\bibinfo {author} {\bibfnamefont {T.}~\bibnamefont
  {Langen}}, \bibinfo {author} {\bibfnamefont {S.}~\bibnamefont {Erne}},
  \bibinfo {author} {\bibfnamefont {R.}~\bibnamefont {Geiger}}, \bibinfo
  {author} {\bibfnamefont {B.}~\bibnamefont {Rauer}}, \bibinfo {author}
  {\bibfnamefont {T.}~\bibnamefont {Schweigler}}, \bibinfo {author}
  {\bibfnamefont {M.}~\bibnamefont {Kuhnert}}, \bibinfo {author} {\bibfnamefont
  {W.}~\bibnamefont {Rohringer}}, \bibinfo {author} {\bibfnamefont {I.~E.}\
  \bibnamefont {Mazets}}, \bibinfo {author} {\bibfnamefont {T.}~\bibnamefont
  {Gasenzer}}, \ and\ \bibinfo {author} {\bibfnamefont {J.}~\bibnamefont
  {Schmiedmayer}},\ }\href {\doibase 10.1126/science.1257026} {\bibfield
  {journal} {\bibinfo  {journal} {Science}\ }\textbf {\bibinfo {volume}
  {348}},\ \bibinfo {pages} {207} (\bibinfo {year} {2015})}\BibitemShut
  {NoStop}%
\bibitem [{\citenamefont {Feng}\ \emph {et~al.}(2019)\citenamefont {Feng},
  \citenamefont {Hu}, \citenamefont {Clark},\ and\ \citenamefont
  {Chin}}]{Feng2019}%
  \BibitemOpen
  \bibfield  {author} {\bibinfo {author} {\bibfnamefont {L.}~\bibnamefont
  {Feng}}, \bibinfo {author} {\bibfnamefont {J.}~\bibnamefont {Hu}}, \bibinfo
  {author} {\bibfnamefont {L.~W.}\ \bibnamefont {Clark}}, \ and\ \bibinfo
  {author} {\bibfnamefont {C.}~\bibnamefont {Chin}},\ }\href {\doibase
  10.1126/science.aat5008} {\bibfield  {journal} {\bibinfo  {journal}
  {Science}\ }\textbf {\bibinfo {volume} {363}},\ \bibinfo {pages} {521}
  (\bibinfo {year} {2019})},\ \Eprint {http://arxiv.org/abs/1803.01786}
  {arXiv:1803.01786} \BibitemShut {NoStop}%
\bibitem [{\citenamefont {Zache}\ \emph {et~al.}(2020)\citenamefont {Zache},
  \citenamefont {Schweigler}, \citenamefont {Erne}, \citenamefont
  {Schmiedmayer},\ and\ \citenamefont {Berges}}]{zache2020extracting}%
  \BibitemOpen
  \bibfield  {author} {\bibinfo {author} {\bibfnamefont {T.~V.}\ \bibnamefont
  {Zache}}, \bibinfo {author} {\bibfnamefont {T.}~\bibnamefont {Schweigler}},
  \bibinfo {author} {\bibfnamefont {S.}~\bibnamefont {Erne}}, \bibinfo {author}
  {\bibfnamefont {J.}~\bibnamefont {Schmiedmayer}}, \ and\ \bibinfo {author}
  {\bibfnamefont {J.}~\bibnamefont {Berges}},\ }\href {\doibase
  https://doi.org/10.1103/PhysRevX.10.011020} {\bibfield  {journal} {\bibinfo
  {journal} {Physical Review X}\ }\textbf {\bibinfo {volume} {10}},\ \bibinfo
  {pages} {011020} (\bibinfo {year} {2020})}\BibitemShut {NoStop}%
\bibitem [{\citenamefont {Pr{\"u}fer}\ \emph {et~al.}(2020)\citenamefont
  {Pr{\"u}fer}, \citenamefont {Zache}, \citenamefont {Kunkel}, \citenamefont
  {Lannig}, \citenamefont {Bonnin}, \citenamefont {Strobel}, \citenamefont
  {Berges},\ and\ \citenamefont {Oberthaler}}]{prufer2020experimental}%
  \BibitemOpen
  \bibfield  {author} {\bibinfo {author} {\bibfnamefont {M.}~\bibnamefont
  {Pr{\"u}fer}}, \bibinfo {author} {\bibfnamefont {T.~V.}\ \bibnamefont
  {Zache}}, \bibinfo {author} {\bibfnamefont {P.}~\bibnamefont {Kunkel}},
  \bibinfo {author} {\bibfnamefont {S.}~\bibnamefont {Lannig}}, \bibinfo
  {author} {\bibfnamefont {A.}~\bibnamefont {Bonnin}}, \bibinfo {author}
  {\bibfnamefont {H.}~\bibnamefont {Strobel}}, \bibinfo {author} {\bibfnamefont
  {J.}~\bibnamefont {Berges}}, \ and\ \bibinfo {author} {\bibfnamefont {M.~K.}\
  \bibnamefont {Oberthaler}},\ }\href {\doibase
  https://doi.org/10.1038/s41567-020-0933-6} {\bibfield  {journal} {\bibinfo
  {journal} {Nature Physics}\ }\textbf {\bibinfo {volume} {16}},\ \bibinfo
  {pages} {1012} (\bibinfo {year} {2020})}\BibitemShut {NoStop}%
\bibitem [{\citenamefont {Schweigler}\ \emph {et~al.}(2021)\citenamefont
  {Schweigler}, \citenamefont {Gluza}, \citenamefont {Tajik}, \citenamefont
  {Sotiriadis}, \citenamefont {Cataldini}, \citenamefont {Ji}, \citenamefont
  {M{\o}ller}, \citenamefont {Sabino}, \citenamefont {Rauer}, \citenamefont
  {Eisert} \emph {et~al.}}]{schweigler2021decay}%
  \BibitemOpen
  \bibfield  {author} {\bibinfo {author} {\bibfnamefont {T.}~\bibnamefont
  {Schweigler}}, \bibinfo {author} {\bibfnamefont {M.}~\bibnamefont {Gluza}},
  \bibinfo {author} {\bibfnamefont {M.}~\bibnamefont {Tajik}}, \bibinfo
  {author} {\bibfnamefont {S.}~\bibnamefont {Sotiriadis}}, \bibinfo {author}
  {\bibfnamefont {F.}~\bibnamefont {Cataldini}}, \bibinfo {author}
  {\bibfnamefont {S.-C.}\ \bibnamefont {Ji}}, \bibinfo {author} {\bibfnamefont
  {F.~S.}\ \bibnamefont {M{\o}ller}}, \bibinfo {author} {\bibfnamefont
  {J.}~\bibnamefont {Sabino}}, \bibinfo {author} {\bibfnamefont
  {B.}~\bibnamefont {Rauer}}, \bibinfo {author} {\bibfnamefont
  {J.}~\bibnamefont {Eisert}},  \emph {et~al.},\ }\href {\doibase
  10.1038/s41567-020-01139-2} {\bibfield  {journal} {\bibinfo  {journal}
  {Nature Physics}\ }\textbf {\bibinfo {volume} {17}},\ \bibinfo {pages} {559}
  (\bibinfo {year} {2021})}\BibitemShut {NoStop}%
\bibitem [{\citenamefont {Beck}\ \emph {et~al.}(2018)\citenamefont {Beck},
  \citenamefont {Mazets},\ and\ \citenamefont
  {Schweigler}}]{SchweiglerTransfer2018}%
  \BibitemOpen
  \bibfield  {author} {\bibinfo {author} {\bibfnamefont {S.}~\bibnamefont
  {Beck}}, \bibinfo {author} {\bibfnamefont {I.~E.}\ \bibnamefont {Mazets}}, \
  and\ \bibinfo {author} {\bibfnamefont {T.}~\bibnamefont {Schweigler}},\
  }\href {\doibase 10.1103/PhysRevA.98.023613} {\bibfield  {journal} {\bibinfo
  {journal} {Phys. Rev. A}\ }\textbf {\bibinfo {volume} {98}},\ \bibinfo
  {pages} {023613} (\bibinfo {year} {2018})}\BibitemShut {NoStop}%
\bibitem [{\citenamefont {Calzetta}\ and\ \citenamefont
  {Hu}(2009)}]{calzetta2009nonequilibrium}%
  \BibitemOpen
  \bibfield  {author} {\bibinfo {author} {\bibfnamefont {E.~A.}\ \bibnamefont
  {Calzetta}}\ and\ \bibinfo {author} {\bibfnamefont {B.-L.~B.}\ \bibnamefont
  {Hu}},\ }\href@noop {} {\emph {\bibinfo {title} {Nonequilibrium quantum field
  theory}}}\ (\bibinfo  {publisher} {Cambridge University Press},\ \bibinfo
  {year} {2009})\BibitemShut {NoStop}%
\bibitem [{\citenamefont {Kunkel}\ \emph {et~al.}(2019)\citenamefont {Kunkel},
  \citenamefont {Pr{\"u}fer}, \citenamefont {Lannig}, \citenamefont
  {Rosa-Medina}, \citenamefont {Bonnin}, \citenamefont {G{\"a}rttner},
  \citenamefont {Strobel},\ and\ \citenamefont
  {Oberthaler}}]{kunkel2019simultaneous}%
  \BibitemOpen
  \bibfield  {author} {\bibinfo {author} {\bibfnamefont {P.}~\bibnamefont
  {Kunkel}}, \bibinfo {author} {\bibfnamefont {M.}~\bibnamefont {Pr{\"u}fer}},
  \bibinfo {author} {\bibfnamefont {S.}~\bibnamefont {Lannig}}, \bibinfo
  {author} {\bibfnamefont {R.}~\bibnamefont {Rosa-Medina}}, \bibinfo {author}
  {\bibfnamefont {A.}~\bibnamefont {Bonnin}}, \bibinfo {author} {\bibfnamefont
  {M.}~\bibnamefont {G{\"a}rttner}}, \bibinfo {author} {\bibfnamefont
  {H.}~\bibnamefont {Strobel}}, \ and\ \bibinfo {author} {\bibfnamefont
  {M.~K.}\ \bibnamefont {Oberthaler}},\ }\href {\doibase
  https://doi.org/10.1103/PhysRevLett.123.063603} {\bibfield  {journal}
  {\bibinfo  {journal} {Physical review letters}\ }\textbf {\bibinfo {volume}
  {123}},\ \bibinfo {pages} {063603} (\bibinfo {year} {2019})}\BibitemShut
  {NoStop}%
\bibitem [{\citenamefont {Gluza}\ \emph {et~al.}(2020)\citenamefont {Gluza},
  \citenamefont {Schweigler}, \citenamefont {Rauer}, \citenamefont {Krumnow},
  \citenamefont {Schmiedmayer},\ and\ \citenamefont
  {Eisert}}]{gluza2020quantum}%
  \BibitemOpen
  \bibfield  {author} {\bibinfo {author} {\bibfnamefont {M.}~\bibnamefont
  {Gluza}}, \bibinfo {author} {\bibfnamefont {T.}~\bibnamefont {Schweigler}},
  \bibinfo {author} {\bibfnamefont {B.}~\bibnamefont {Rauer}}, \bibinfo
  {author} {\bibfnamefont {C.}~\bibnamefont {Krumnow}}, \bibinfo {author}
  {\bibfnamefont {J.}~\bibnamefont {Schmiedmayer}}, \ and\ \bibinfo {author}
  {\bibfnamefont {J.}~\bibnamefont {Eisert}},\ }\href {\doibase
  https://doi.org/10.1038/s42005-019-0273-y} {\bibfield  {journal} {\bibinfo
  {journal} {Communications Physics}\ }\textbf {\bibinfo {volume} {3}},\
  \bibinfo {pages} {12} (\bibinfo {year} {2020})}\BibitemShut {NoStop}%
\bibitem [{\citenamefont {Cooper}\ \emph {et~al.}(2022)\citenamefont {Cooper},
  \citenamefont {Kunkel}, \citenamefont {Periwal},\ and\ \citenamefont
  {Schleier-Smith}}]{cooper2022engineering}%
  \BibitemOpen
  \bibfield  {author} {\bibinfo {author} {\bibfnamefont {E.~S.}\ \bibnamefont
  {Cooper}}, \bibinfo {author} {\bibfnamefont {P.}~\bibnamefont {Kunkel}},
  \bibinfo {author} {\bibfnamefont {A.}~\bibnamefont {Periwal}}, \ and\
  \bibinfo {author} {\bibfnamefont {M.}~\bibnamefont {Schleier-Smith}},\ }\href
  {\doibase https://doi.org/10.48550/arXiv.2212.11961} {\bibfield  {journal}
  {\bibinfo  {journal} {arXiv preprint arXiv:2212.11961}\ } (\bibinfo {year}
  {2022}),\ https://doi.org/10.48550/arXiv.2212.11961}\BibitemShut {NoStop}%
\bibitem [{\citenamefont {Elben}\ \emph {et~al.}(2023)\citenamefont {Elben},
  \citenamefont {Flammia}, \citenamefont {Huang}, \citenamefont {Kueng},
  \citenamefont {Preskill}, \citenamefont {Vermersch},\ and\ \citenamefont
  {Zoller}}]{elben2023randomized}%
  \BibitemOpen
  \bibfield  {author} {\bibinfo {author} {\bibfnamefont {A.}~\bibnamefont
  {Elben}}, \bibinfo {author} {\bibfnamefont {S.~T.}\ \bibnamefont {Flammia}},
  \bibinfo {author} {\bibfnamefont {H.-Y.}\ \bibnamefont {Huang}}, \bibinfo
  {author} {\bibfnamefont {R.}~\bibnamefont {Kueng}}, \bibinfo {author}
  {\bibfnamefont {J.}~\bibnamefont {Preskill}}, \bibinfo {author}
  {\bibfnamefont {B.}~\bibnamefont {Vermersch}}, \ and\ \bibinfo {author}
  {\bibfnamefont {P.}~\bibnamefont {Zoller}},\ }\href {\doibase
  https://doi.org/10.1038/s42254-022-00535-2} {\bibfield  {journal} {\bibinfo
  {journal} {Nature Reviews Physics}\ }\textbf {\bibinfo {volume} {5}},\
  \bibinfo {pages} {9} (\bibinfo {year} {2023})}\BibitemShut {NoStop}%
\bibitem [{\citenamefont {Huang}\ \emph {et~al.}(2020)\citenamefont {Huang},
  \citenamefont {Kueng},\ and\ \citenamefont {Preskill}}]{huang2020predicting}%
  \BibitemOpen
  \bibfield  {author} {\bibinfo {author} {\bibfnamefont {H.-Y.}\ \bibnamefont
  {Huang}}, \bibinfo {author} {\bibfnamefont {R.}~\bibnamefont {Kueng}}, \ and\
  \bibinfo {author} {\bibfnamefont {J.}~\bibnamefont {Preskill}},\ }\href
  {\doibase https://doi.org/10.1038/s41567-020-0932-7} {\bibfield  {journal}
  {\bibinfo  {journal} {Nature Physics}\ }\textbf {\bibinfo {volume} {16}},\
  \bibinfo {pages} {1050} (\bibinfo {year} {2020})}\BibitemShut {NoStop}%
\bibitem [{\citenamefont {Bairey}\ \emph {et~al.}(2020)\citenamefont {Bairey},
  \citenamefont {Guo}, \citenamefont {Poletti}, \citenamefont {Lindner},\ and\
  \citenamefont {Arad}}]{bairey2020learning}%
  \BibitemOpen
  \bibfield  {author} {\bibinfo {author} {\bibfnamefont {E.}~\bibnamefont
  {Bairey}}, \bibinfo {author} {\bibfnamefont {C.}~\bibnamefont {Guo}},
  \bibinfo {author} {\bibfnamefont {D.}~\bibnamefont {Poletti}}, \bibinfo
  {author} {\bibfnamefont {N.~H.}\ \bibnamefont {Lindner}}, \ and\ \bibinfo
  {author} {\bibfnamefont {I.}~\bibnamefont {Arad}},\ }\href {\doibase
  10.1088/1367-2630/ab73cd} {\bibfield  {journal} {\bibinfo  {journal} {New
  Journal of Physics}\ }\textbf {\bibinfo {volume} {22}},\ \bibinfo {pages}
  {032001} (\bibinfo {year} {2020})}\BibitemShut {NoStop}%
\bibitem [{\citenamefont {Kokail}\ \emph {et~al.}(2021)\citenamefont {Kokail},
  \citenamefont {van Bijnen}, \citenamefont {Elben}, \citenamefont
  {Vermersch},\ and\ \citenamefont {Zoller}}]{kokail2021entanglement}%
  \BibitemOpen
  \bibfield  {author} {\bibinfo {author} {\bibfnamefont {C.}~\bibnamefont
  {Kokail}}, \bibinfo {author} {\bibfnamefont {R.}~\bibnamefont {van Bijnen}},
  \bibinfo {author} {\bibfnamefont {A.}~\bibnamefont {Elben}}, \bibinfo
  {author} {\bibfnamefont {B.}~\bibnamefont {Vermersch}}, \ and\ \bibinfo
  {author} {\bibfnamefont {P.}~\bibnamefont {Zoller}},\ }\href@noop {}
  {\bibfield  {journal} {\bibinfo  {journal} {Nature Physics}\ }\textbf
  {\bibinfo {volume} {17}},\ \bibinfo {pages} {936} (\bibinfo {year}
  {2021})}\BibitemShut {NoStop}%
\bibitem [{\citenamefont {Joshi}\ \emph {et~al.}(2023)\citenamefont {Joshi},
  \citenamefont {Kokail}, \citenamefont {van Bijnen}, \citenamefont {Kranzl},
  \citenamefont {Zache}, \citenamefont {Blatt}, \citenamefont {Roos},\ and\
  \citenamefont {Zoller}}]{joshi2023exploring}%
  \BibitemOpen
  \bibfield  {author} {\bibinfo {author} {\bibfnamefont {M.~K.}\ \bibnamefont
  {Joshi}}, \bibinfo {author} {\bibfnamefont {C.}~\bibnamefont {Kokail}},
  \bibinfo {author} {\bibfnamefont {R.}~\bibnamefont {van Bijnen}}, \bibinfo
  {author} {\bibfnamefont {F.}~\bibnamefont {Kranzl}}, \bibinfo {author}
  {\bibfnamefont {T.~V.}\ \bibnamefont {Zache}}, \bibinfo {author}
  {\bibfnamefont {R.}~\bibnamefont {Blatt}}, \bibinfo {author} {\bibfnamefont
  {C.~F.}\ \bibnamefont {Roos}}, \ and\ \bibinfo {author} {\bibfnamefont
  {P.}~\bibnamefont {Zoller}},\ }\href {\doibase
  https://doi.org/10.1038/s41586-023-06768-0} {\bibfield  {journal} {\bibinfo
  {journal} {Nature}\ }\textbf {\bibinfo {volume} {624}},\ \bibinfo {pages}
  {539} (\bibinfo {year} {2023})}\BibitemShut {NoStop}%
\bibitem [{\citenamefont {Calabrese}\ and\ \citenamefont
  {Cardy}(2004)}]{calabrese2004entanglement}%
  \BibitemOpen
  \bibfield  {author} {\bibinfo {author} {\bibfnamefont {P.}~\bibnamefont
  {Calabrese}}\ and\ \bibinfo {author} {\bibfnamefont {J.}~\bibnamefont
  {Cardy}},\ }\href {\doibase 10.1088/1742-5468/2004/06/P06002} {\bibfield
  {journal} {\bibinfo  {journal} {Journal of statistical mechanics: theory and
  experiment}\ }\textbf {\bibinfo {volume} {2004}},\ \bibinfo {pages} {P06002}
  (\bibinfo {year} {2004})}\BibitemShut {NoStop}%
\bibitem [{\citenamefont {Fr{\"o}lian}\ \emph {et~al.}(2022)\citenamefont
  {Fr{\"o}lian}, \citenamefont {Chisholm}, \citenamefont {Neri}, \citenamefont
  {Cabrera}, \citenamefont {Ramos}, \citenamefont {Celi},\ and\ \citenamefont
  {Tarruell}}]{frolian2022realizing}%
  \BibitemOpen
  \bibfield  {author} {\bibinfo {author} {\bibfnamefont {A.}~\bibnamefont
  {Fr{\"o}lian}}, \bibinfo {author} {\bibfnamefont {C.~S.}\ \bibnamefont
  {Chisholm}}, \bibinfo {author} {\bibfnamefont {E.}~\bibnamefont {Neri}},
  \bibinfo {author} {\bibfnamefont {C.~R.}\ \bibnamefont {Cabrera}}, \bibinfo
  {author} {\bibfnamefont {R.}~\bibnamefont {Ramos}}, \bibinfo {author}
  {\bibfnamefont {A.}~\bibnamefont {Celi}}, \ and\ \bibinfo {author}
  {\bibfnamefont {L.}~\bibnamefont {Tarruell}},\ }\href {\doibase
  https://doi.org/10.1038/s41586-022-04943-3} {\bibfield  {journal} {\bibinfo
  {journal} {Nature}\ }\textbf {\bibinfo {volume} {608}},\ \bibinfo {pages}
  {293} (\bibinfo {year} {2022})}\BibitemShut {NoStop}%
\bibitem [{\citenamefont {Semeghini}\ \emph {et~al.}(2021)\citenamefont
  {Semeghini}, \citenamefont {Levine}, \citenamefont {Keesling}, \citenamefont
  {Ebadi}, \citenamefont {Wang}, \citenamefont {Bluvstein}, \citenamefont
  {Verresen}, \citenamefont {Pichler}, \citenamefont {Kalinowski},
  \citenamefont {Samajdar} \emph {et~al.}}]{semeghini2021probing}%
  \BibitemOpen
  \bibfield  {author} {\bibinfo {author} {\bibfnamefont {G.}~\bibnamefont
  {Semeghini}}, \bibinfo {author} {\bibfnamefont {H.}~\bibnamefont {Levine}},
  \bibinfo {author} {\bibfnamefont {A.}~\bibnamefont {Keesling}}, \bibinfo
  {author} {\bibfnamefont {S.}~\bibnamefont {Ebadi}}, \bibinfo {author}
  {\bibfnamefont {T.~T.}\ \bibnamefont {Wang}}, \bibinfo {author}
  {\bibfnamefont {D.}~\bibnamefont {Bluvstein}}, \bibinfo {author}
  {\bibfnamefont {R.}~\bibnamefont {Verresen}}, \bibinfo {author}
  {\bibfnamefont {H.}~\bibnamefont {Pichler}}, \bibinfo {author} {\bibfnamefont
  {M.}~\bibnamefont {Kalinowski}}, \bibinfo {author} {\bibfnamefont
  {R.}~\bibnamefont {Samajdar}},  \emph {et~al.},\ }\href@noop {} {\bibfield
  {journal} {\bibinfo  {journal} {Science}\ }\textbf {\bibinfo {volume}
  {374}},\ \bibinfo {pages} {1242} (\bibinfo {year} {2021})}\BibitemShut
  {NoStop}%
\end{thebibliography}%

\end{document}